# IOTA (Integrable Optics Test Accelerator): Facility and Experimental Beam Physics Program

**Sergei Antipov, Daniel Broemmelsiek, David Bruhwiler\*, Dean Edstrom, Elvin Harms, Valery Lebedev, Jerry Leibfritz, Sergei Nagaitsev, Chong-Shik Park, Henryk Piekarz, Philippe Piot\*\*, Eric Prebys, Alexander Romanov, Jinhao Ruan, Tanaji Sen, Giulio Stancari, Charles Thangaraj, Randy Thurman-Keup, Alexander Valishev, Vladimir Shiltsev\*\*\***

*Fermi National Accelerator Laboratory, Batavia, Illinois 60510, USA*
*\*RadiaSoft LLC, Boulder, Colorado 80304, USA*
*\*\* also at Northern Illinois University, DeKalb, Ilinois, 60115, USA*

*\*\*\*E-mail*: shiltsev@fnal.gov

ABSTRACT: The Integrable Optics Test Accelerator (IOTA) is a storage ring for advanced beam physics research currently being built and commissioned at Fermilab. It will operate with protons and electrons using injectors with momenta of 70 and 150 MeV/c, respectively. The research program includes the study of nonlinear focusing integrable optical beam lattices based on special magnets and electron lenses, beam dynamics of space-charge effects and their compensation, optical stochastic cooling, and several other experiments. In this article, we present the design and main parameters of the facility, outline progress to date and provide the timeline of the construction, commissioning and research. The physical principles, design, and hardware implementation plans for the major IOTA experiments are also discussed.

# Contents

## 1. Introduction

### 1.1 US high energy physics plan and motivation for IOTA

The 2014 Particle Physics Project Prioritization Panel (P5) report [1] identified a high energy neutrino program to determine the mass hierarchy and measure CP violation as the top priority of the US domestic intensity frontier high-energy physics for the next 20-30 years. The program will be based on the Fermilab accelerator complex, which needs to be upgraded for increased proton intensity. To this end, a new beam line - the Long Baseline Neutrino Facility (LBNF), new experiment - the Deep Underground Neutrino Experiment (DUNE), located in the Sanford Underground Research Facility (SURF) [2], and new accelerator - the PIP-II (Proton Improvement Plan – II) superconducting RF linac [3], are being planned.

The P5 physics goals require about 900 kt·MW·years of total exposure (product of the neutrino detector mass, average proton beam power on the neutrino target and data taking period). This can be achieved assuming a 40 kton Liquid Argon detector and accelerator operation with the eventual multi-MW beam power of LBNF. Recently, "6+6 batch slip-stacking" was commissioned in the Fermilab Recycler, which reduced from the 2.2s it was during the MINOS/Collider Run II era to 1.33s. This resulted in a world-record 615 kW average proton beam power over one hour to the NuMI beam line. Sustainable routine operation at the 700kW level is expected after an upgrade of the Recycler beam collimation system [4].

Construction of a new SRF 800 MeV H-/proton linac as part of the PIP-II project is expected to increase the Booster per pulse intensity by 50% and allow delivery of 1.2 MW of beam power at 120 GeV from Fermilab's Main Injector, with power approaching 1 MW at energies as low as 60 GeV [5], at the start of DUNE/LBNF operations, ca 2025.

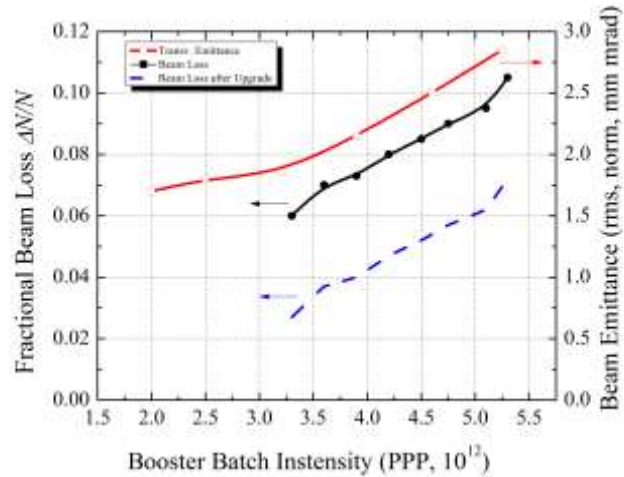

**Figure 1:** Fermilab's 8 GeV proton Booster synchrotron performance vs intensity: beam emittance (red line, right axis) and fractional beam loss at present (black, left axis) and after currently planned upgrades (blue). Intensity dependent effects lead to unacceptable losses and emittance growth (from [6], data courtesy W.Pellico).

Further progress of the intensity frontier accelerators is hindered by fundamental beam physics phenomena such as space-charge effects (Fig. 1), beam halo formation, particle losses, transverse and longitudinal instabilities, beam loading, inefficiencies of beam injection and extraction, etc. An extensive accelerator R&D program towards multi-MW beams has been started at Fermilab. Besides theoretical investigations, there have been modeling and simulation efforts dedicated to understanding beam dynamics issues with high intensity beams in the existing accelerators. There are also three important experimental activities: demonstration of novel techniques for high-current beam accelerators at the Integrable Optics Test Accelerator (IOTA), development of cost-effective SC RF cavities and development of high-power targets capable to accept muliti-MW beams [6]. In this article we present the design and main parameters of the IOTA facility, outline progress to date and the timeline of the construction, commissioning and research, and describe the physical principles, design, and hardware implementation plans for the IOTA experiments.

**1.2 Main requirements and parameters of IOTA**

The Integrable Optics Test Accelerator (IOTA) at Fermilab's Accelerator Science and Technology (FAST) facility [7] is a storage ring for advanced high intensity beam physics research. The IOTA research program includes the study of nonlinear focusing integrable optical beam lattices based on special magnets and electron lenses, beam dynamics of ultimate space-charge effects and their compensation, optical stochastic cooling, and several other experiments – see Section 3 below. A low energy proton or H- injector is needed to study space-charge dominated circulating beams in relatively compact storage ring. Some experiments, such as the integrable optics test require very tight control of the accelerator focusing lattice – e.g., to better than 1% percent accuracy in $\beta$-functions – that cannot be achieved with protons, and therefore, require a pencil beam of modestly relativistic electrons free of space-charge complications. Adequate control and flexibility of the beam orbits and lattice and advanced, large dynamic range diagnostics of all important parameters of both proton and electron beams (intensity, halo, emittance, bunch length, energy spread, transverse and longitudinal intensity profiles, etc.) are absolutely critical for the IOTA R&D program. Table 1 presents key required parameters of the IOTA/FAST facility and Fig. 2 schematically shows its layout. The experimental accelerator research program, augmented with corresponding modeling and design efforts, has been started at Fermilab in collaboration more than two dozen partners, including universities, SBIR industrial companies, US National Laboratories, and international partners [8].

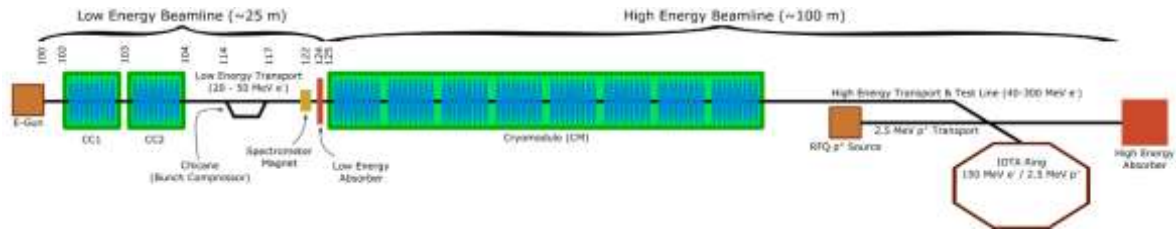

**Figure 2:** Schematic layout of the IOTA ring and its two injectors. The facility includes (from left to right): normal conducting 5 MeV electron RF photoinjector, two 1.3 SRF booster cavities CC1 and CC2, magnetic chicane, 12 m long 1.3 SRF cryomodule, high energy electron transport beamline, 2.5 MeV RFQ proton injector, IOTA ring, and high energy electron beam absorber.

**Table 1:** Key design parameters of the IOTA ring and its injectors

| Parameter | Value | |
|---|---|---|
| Circumference | 40 m | |
| Experimental straight sections | 4 | |
| Particle species | protons/*H*- ions | electrons |
| Particle momentum (max) | 70 MeV/c | 150 MeV/c |
| Maximum space-charge parameter $\Delta Q_{sc}$ | ≥ 0.5 | ~0 |

**2 Design of IOTA and its Injectors**

**2.1 IOTA ring design**

The IOTA ring design is determined by the demands of the experimental program, limitations of the available space and cost optimization. The main requirements to the ring design are the following:
- capability of circulating either proton (up to 70 MeV/c momentum or 2.5 MeV kinetic energy) or electron (up to 150 MeV) beams;
- significantly large beam pipe aperture to accommodate large-amplitude oscillations of pencil beams and large size proton beam;

- significantly long straight sections to accommodate the experimental apparatus, including the nonlinear insert(s), and small enough footprint to fit in the existing FAST machine hall;
- significant flexibility of the optical lattice to accommodate all experimental options;
- precise control of the beam optics quality and stability;
- cost-effective solution based on conventional technology (magnets, RF).

The IOTA ring parameters are listed in Tables 2 and 3 and its layout is depicted in Fig. 3, where the major magnetic and diagnostic elements are shown.

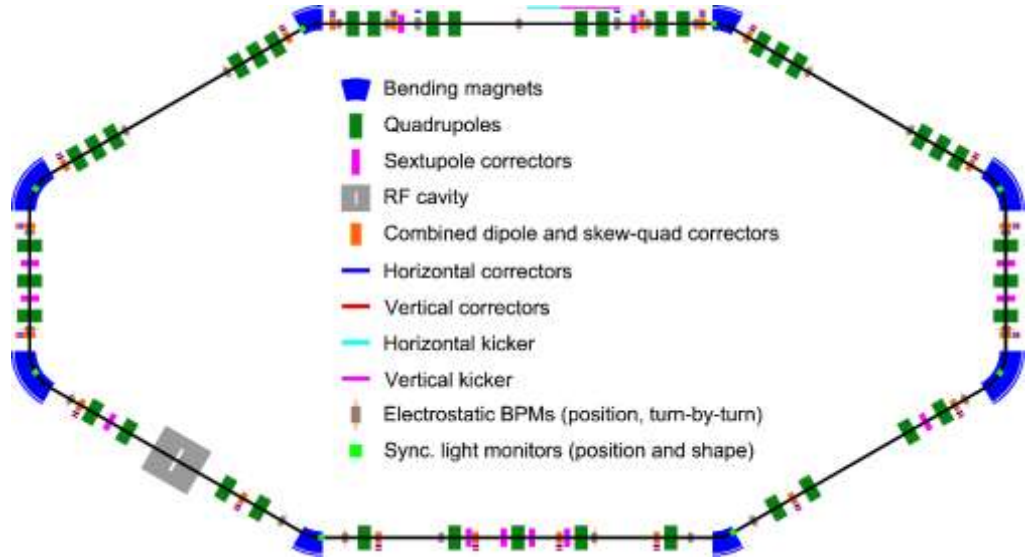

**Figure 3**: Layout of the Integrable Optics Test Accelerator (IOTA) ring.

The ring geometry is defined by 8 main bending dipole magnets (4 30-degree and 4 60-degree bends) and 6 long and 2 short straight sections. The 2 short sections between 60-degree magnets, shown vertically in the figure, are used for dispersion suppression and chromaticity correction. The top horizontal section is used for the beam injection. The lower-left diagonal section is occupied by the accelerating RF cavity. The remaining 4 long sections are designated for the installation of experimental apparatus, such as the nonlinear inserts for integrable optics, the electron lens, and optical stochastic cooling. The ring circumference is 40 m. The nominal bending magnetic field of 0.7 T allows for the maximum electron beam energy of 150 MeV. The 2.5 MeV kinetic energy of the proton beam from the proton injector corresponds to the momentum of 70 MeV/c and requires half the nominal bending field.

Focusing is provided by 39 quadrupole magnets. The correction system consists of 20 combined-function dipole (horizontal and vertical) and skew-quadrupole corrector magnets, 8 horizontal orbit correction coils incorporated in the main dipoles, 2 special vertical correctors in the injection straight section, and 10 sextupoles for chromaticity correction. The lattice is mirror-symmetrical with respect to the vertical center line, allowing for a cost-efficient powering scheme with 20 power supplies powering the main quadrupole magnets (7 250 A, 7 120 A, and 6 70 A supplies). The main dipoles are powered in series with the injection Lambertson by a single power supply. All of the corrector elements are powered by individual bipolar 2 A/15 V supplies reused from the Tevatron Collider.

The vacuum chamber in the straight sections is made of stainless steel pipe with a 2-inch internal diameter. In bending magnets, the vacuum chamber is Aluminium of rectangular cross-section with dimensions of 50×50 mm. Vacuum at a level of $6\times10^{-10}$ Torr or better is maintained with the use of combination NEG and ion pumping at 35 locations around the ring and in the transfer line. The IOTA ring shares a common vacuum with the FAST injector, where the vacuum specification is $1.5\times10^{-8}$ Torr, and thus the transfer line will employ three 150 L/s combination NEG/ion pumps for differential pumping. The system is bakeable in situ up to 120 °C prior to use and each time the vacuum system is opened for modifications or maintenance. The high quality of vacuum system is necessary for proton beam operation and the vacuum of $6\times10^{-10}$ Torr determines the beam lifetime of 5 min for 2.5 MeV protons. The requirements for electron beam operation are much more relaxed, and a level of $3\times10^{-8}$ Torr is sufficient for a beam lifetime of 30 min of 150 MeV electrons.

The injection system (located in the upper straight section in Fig. 3) consists of a Lambertson magnet and an adjacent stripline kicker (Fig. 4). The horizontal 30°-bend Lambertron magnet delivers injected beam at a 1° vertical angle 27 mm below the ring main plane. The beam is then kicked on to the central orbit with a 25-kV pulse of the 1 m long vertical kicker. The kicker pulse length is suitable for single-turn injection (120 ns revolution period for the electron beam). The injection scheme also requires the localized distortion of the closed orbit, which is achieved by the special vertical orbit correctors up- and downstream of the injection region.

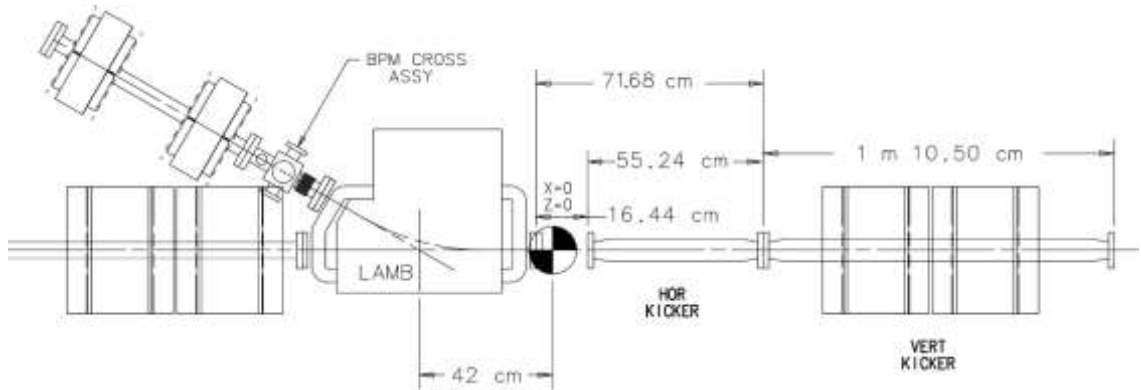

**Figure 4:** Schematic of the IOTA injection region.

The RF accelerating system is provided by a dual-frequency cavity located in the lower-left diagonal section (Fig. 5). The two gaps of the cavity operate at the frequencies of 30 MHz and 2.18 MHz, which correspond to the fourth revolution harmonic for 150 MeV electrons and 2.5 MeV protons, respectively. The high frequency gap can provide up to 1 kV accelerating voltage, sufficient for bunching the 150 MeV electron beam and replenishing the synchrotron radiation losses. The ferrite-loaded cavity has the $Q$ value of 100 and is driven by a 100 W solid-state amplifier. The cavity tuning is achieved by the variable capacitors driven by stepping motors.

The large number of quadrupoles allows for the flexibility of optics tuning to address the needs of beam physics experiments. The current set of optics solutions includes the following options: a) integrable optics (IO) with one nonlinear insert; b) IO with two nonlinear inserts; c) electron lens; d) McMillan electron lens; and e) optical stochastic cooling (OSC). The basic lattice requirements for these options come from the nonlinear integrable optics theory (see Section 3) and include the specific betatron phase advance between the nonlinear insert sections and the zero dispersion in the nonlinear insert. The sampling of wide range nonlinearity requires kicking the probe beam to large betatron amplitudes. Coupled with the machine aperture limitation, this determines the limit on the maximum value of the beta-function. Lastly, the optical stochastic cooling experiment requires a transverse beam emittance of less than 0.1 $\mu$m (rms normalized).

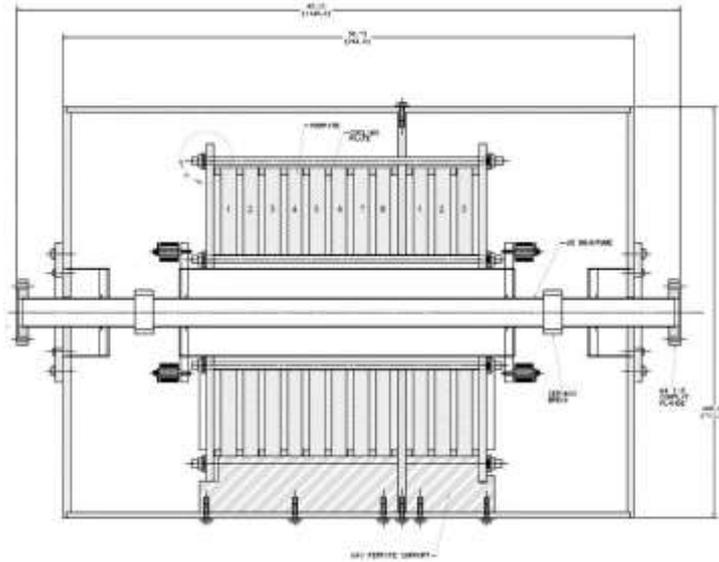

**Figure 5:** Schematic of the IOTA RF cavity. The low-frequency proton cavity is in the left section, high-frequency electron section is on the right.

An optics solution for the IO experiment with a single nonlinear insert is presented in Figs. 6 and 7. This experiment requires a unit transport matrix from one end of the nonlinear insertion to another. The betatron phase advance across the nonlinear insertion must be 0.3 with minimum beta function in the middle for both planes. The length of insertion is 1.8 meters. The amplitude of the pencil beam kick must be 0.4 cm in the horizonal plane and 0.55 cm in vertical. Finally, the required large amplitude of betatron oscillations sets the limit on the ring momentum compaction that should not be smaller than 0.05: the non-isochronous nature of the betatron motion causes phase oscillations for the kicked beam, and the magnitude of these synchrotron oscillations is inversely proportional to the momentum compaction.

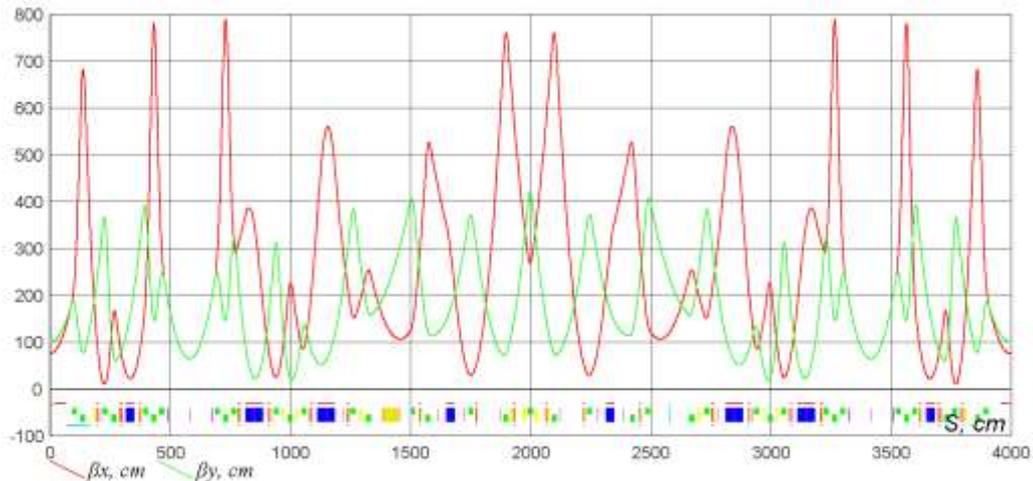

**Figure 6:** IOTA beta-functions for the IO experiment with one nonlinear insert. The origin is at the injection point. The nonlinear insert is located at $s$=500 cm.

The IO experiments require a high precision of the linear optics tuning: the betatron phase advance in the ring must be controlled at the level of $10^{-3}$ (in units of $2\pi$), and the beta-function in the nonlinear insert must be controlled at the level of 1%. For this, a comprehensive set of beam instrumentation (see Section 2.4) and software tools based on the LOCO algorithm [9] have been devised. The simulations, including a realistic set of machine imperfections such as the alignment errors, gradient errors, beam position monitor noise, etc., predict that the required level of precision can be reached.

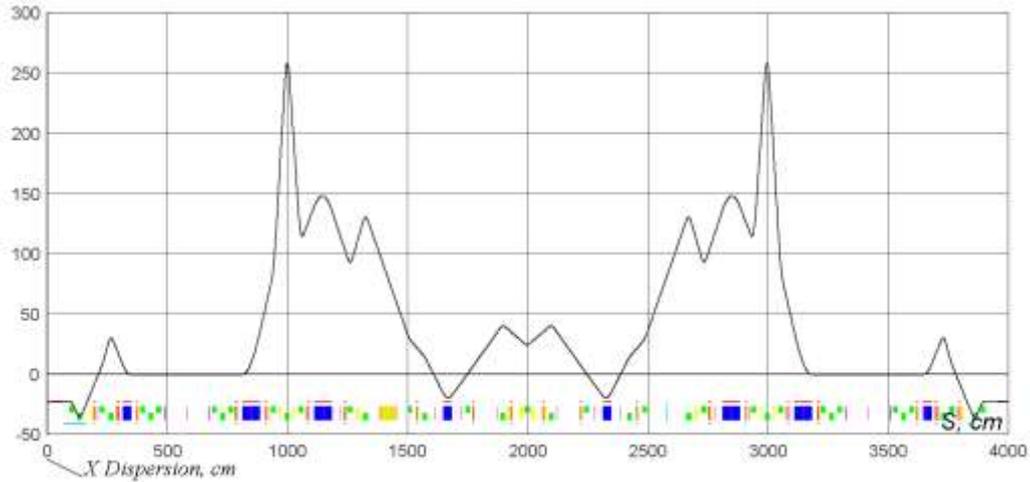

**Figure 7**: IOTA dispersion for the IO experiment with one nonlinear insert. The origin is at the injection point.

The betatron tune chromaticity correction scheme employs 12 sextupole magnets in 6 families that are properly spaced in betatron phase advance to minimize their effect on the dynamics in the different operational scenarios.

**Table 2**: Parameters of the IOTA ring for electron beam operation.

| Parameter | Value |
|---|---|
| Beam energy, $E$ | 150 MeV |
| Circumference, revolution period, $C_0$, $T_0$ | 39.97 m, 133.3 ns |
| Betatron tunes, $Q_x$, $Q_y$ | 4-6 |
| Maximum beta-function, $\beta_x$, $\beta_y$ | 8.5, 4 m |
| Momentum compaction, $\alpha_p$ | 0.067 |
| RF voltage, frequency, revolution harmonic | 1 kV, 30.0 MHz, 4 |
| Synchrotron tune, $Q_s$ | $5.3 \times 10^{-4}$ |
| Number of particles, beam current, $N_e$, $I_e$ | $2 \times 10^9$, 2.4 mA |
| Equilibrium beam emittance, $\varepsilon_x$, $\varepsilon_y$ | 0.04, 0.04 $\mu$m |
| Beam energy spread, bunch length, $\sigma_E$, $\sigma_z$ | $1.35 \times 10^{-4}$, 10.8 cm |
| Radiation damping times, $\tau_x$, $\tau_y$, $\tau_z$ | 0.9 s, 0.9 s, 0.24 s |

**Table 3**: Parameters of the IOTA ring for proton beam operation.

| Parameter | Value |
|---|---|
| Beam kinetic energy, $E$ | 2.5 MeV |
| Revolution period, $C_0$, $T_0$ | 39.97 m, 1.83 $\mu$s |
| RF voltage, frequency, revolution harmonic | 400 V, 2.18 MHz, 4 |
| Synchrotron tune, $Q_s$ | $7 \times 10^{-3}$ |
| Number of particles, beam current, $N_e$, $I_e$ | $9 \times 10^{10}$, 8 mA |
| Equilibrium beam emittance, $\varepsilon_x$, $\varepsilon_y$ | 0.3, 0.3 $\mu$m |
| Beam energy spread, bunch length, $\sigma_E$, $\sigma_z$ | $1.5 \times 10^{-3}$, 1.7 m |
| Space charge tune shift (unbunched, bunched) | -0.5, -1.2 |

## 2.2 IOTA electron injector

The IOTA electron injector comprises a number of components, including a 5 MeV electron RF photoinjector, a 25-meter-long low energy (≤50 MeV) beamline and a ~100-meter-long high energy (≤300 MeV) beamline. The optical functions of the injector are shown in Fig. 8.

The electrons are produced with a UV drive laser. The drive laser is based on a Calmar seed laser, consisting of a Yb-doped fiber laser oscillator running at 1.3 GHz that is then divided down to 81.25 MHz before amplification through a set of fiber amplifiers. The seed output of 81.25 MHz packets of 1054 nm infrared laser is then reduced to the desired pulse train frequency (nominally 3 MHz) with a Pockels cell, before selection of the desired pulse train with two additional Pockels cells and amplification through a series of YLF crystal-based single pass amplifiers (SPAs) and a Northrup-Grumman amplifier. All of this nominally yields 50 µJ of IR per pulse before two frequency-doubling crystal stages generate the green and then the final UV components with a total nominal efficiency of 10% [10]. The pulse train structure is selectable between a single pulse per IOTA beam cycle (nominally 1 Hz) and up to a 5 Hz, 1 ms long train of 3000 pulses at the nominal 3 MHz pulse train frequency, as shown in Fig.9. The UV (263 nm) drive laser pulse train is used to generate an electron pulse in the IOTA electron injector, which is an SRF- based linear accelerator.

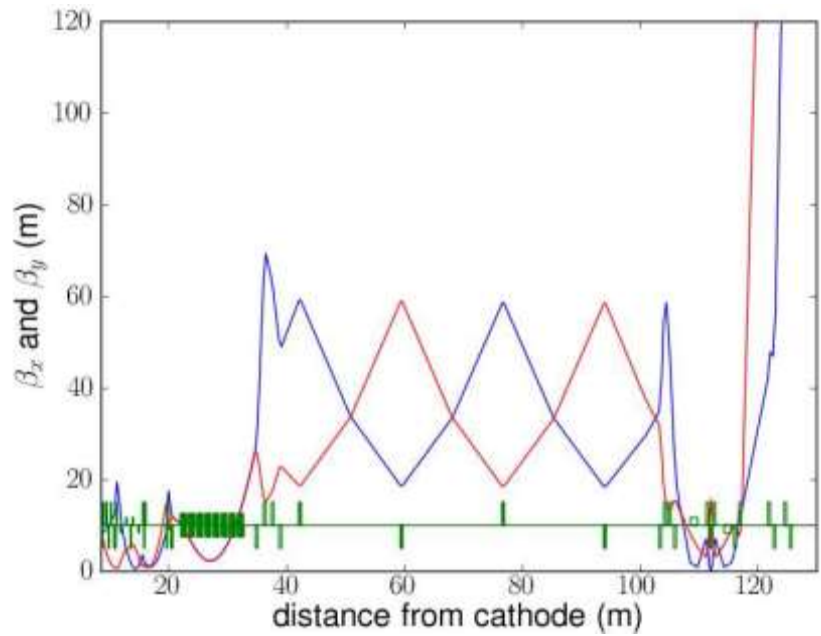

**Figure 8**: Optical functions of the 300 MeV IOTA electron injector linear accelerator. The origin is at the 5 MeV photo-injector cathode, and the beam ends at the high energy absorber.

The 5 MeV electron RF photoinjector is shown in Fig.10. The RF photocathode electron gun is identical to the guns recently developed at DESY Zeuthen (PITZ) for the FLASH facility [11]. It is a normal conducting 1½ cell 1.3 GHz gun operated in the TM010 π mode, with a QL of 11,700, driven by a 5 MW klystron. The power is coupled into the gun via a coaxial RF coupler at the downstream end. The gun is capable of an average DC power dissipation of 20 kW. A temperature feedback system will regulate cooling water temperature to less than ±0.02 °C, as required for good phase stability. The gun can be routinely operated at peak gradients of 40-45 MV/m, with output beam kinetic energy of 4.5 MeV, in a train of electron bunches with up to 4 nC/bunch [12]. The photocathode is a 10 mm diameter molybdenum disk coated with $Cs_2Te$ with 5 mm diameter photosensitive area. The 263 nm wavelength laser light is directed onto the photocathode by a 45° off-axis mirror downstream of the RF coupler. The photocathodes are coated at a separate facility on the Fermilab site, transported under vacuum to the photocathode transfer chamber mounted on the upstream end of the gun, and inserted into the upstream end of the gun via external manipulators, all under vacuum. Several photocathodes have already been prepared and their quantum efficiency measured to be 4-5%. For emittance compensation, the gun is surrounded by 2 solenoid magnets, each with a peak field of 0.28 T at 500 A. Normally the magnet currents are set so the field at the photocathode is 0 in order to minimize the beam emittance; however, the field can also be set to > 1kG at the photocathode for the production of angular-momentum dominated beams and flat beam production. There are four primary types of diagnostics in the injection section: BPMs to measure beam position, transverse profile monitors to measure beam size, resistive wall current monitors to measure beam current and loss monitors to measure beam losses and serve as the primary protection element in the machine protection system. These diagnostics are further discussed in Section 2.4 below.

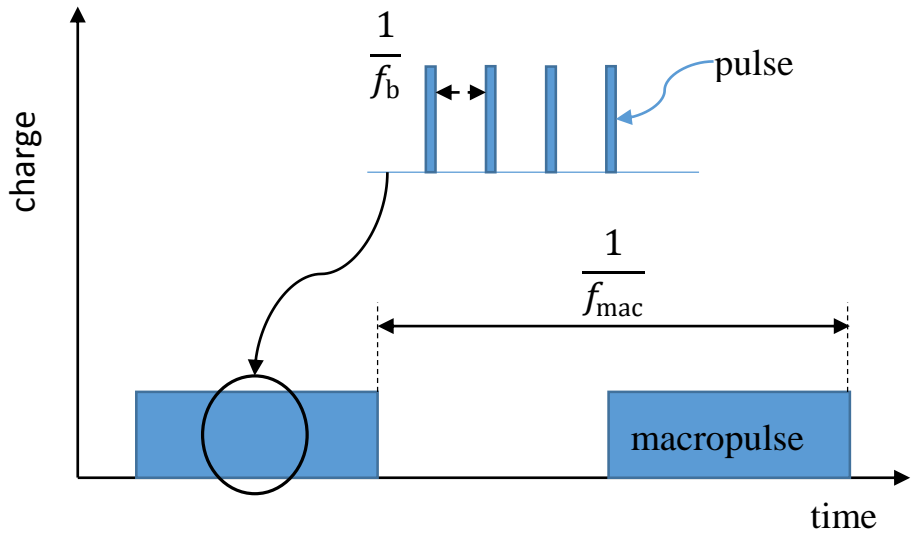

**Figure 9:** Time structure of the IOTA/FAST electron injector. $f_{mac}$ is nominally 1-5 Hz, macropulse length is 1ms, and $f_b$ is 3 MHz.

The electron beam passes through a short (~1 m) diagnostic section before acceleration in two consecutive superconducting RF structures: capture cavities 1 and 2 (CC1 and CC2 in Fig. 11). Each capture cavity is a 9-cell, 1.3 GHz, Nb accelerating structure cooled nominally to 2 K, and each has its own cryostat. Following acceleration of up to 50 MeV, the electron beam passes through the low-energy beam transport section, which includes steering and focusing elements, and an optional chicane ($R_{56}$ = -0.18 m) for bunch compression and beam transforms. A spectrometer dipole magnet, when energized, directs the beam into an absorber prior to the high energy section. The absorber is capable of absorbing up to 400 W of beam power.

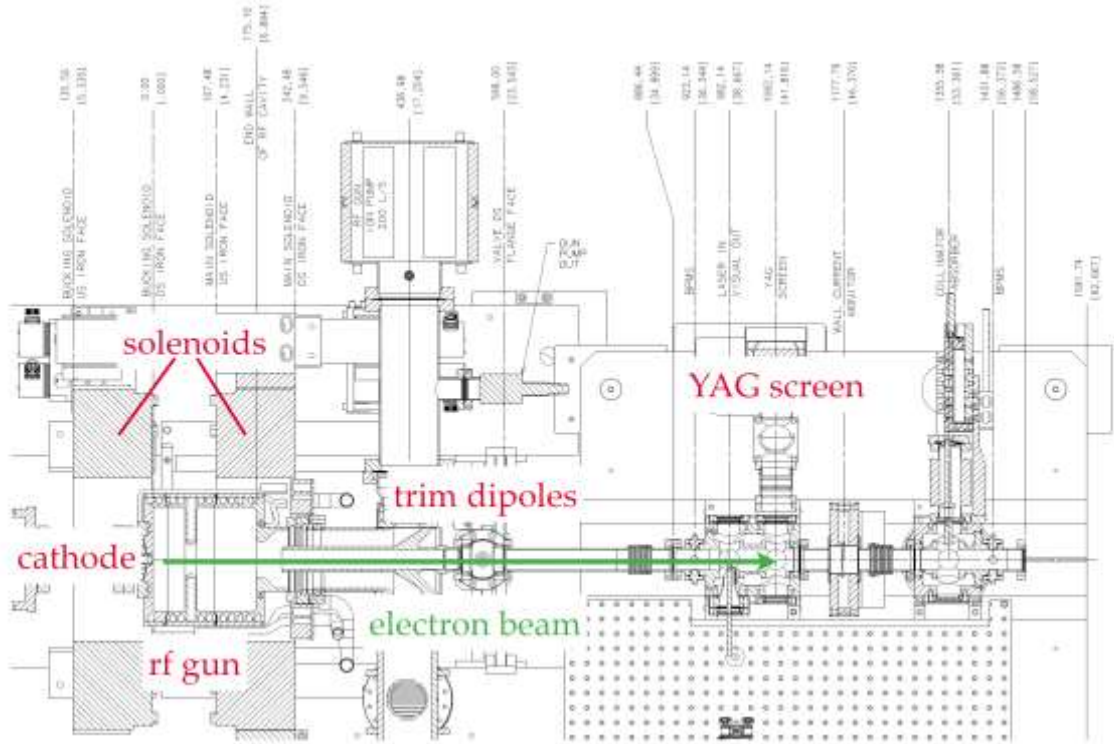

**Figure 10:** Schematic layout of IOTA/FAST electron injector.

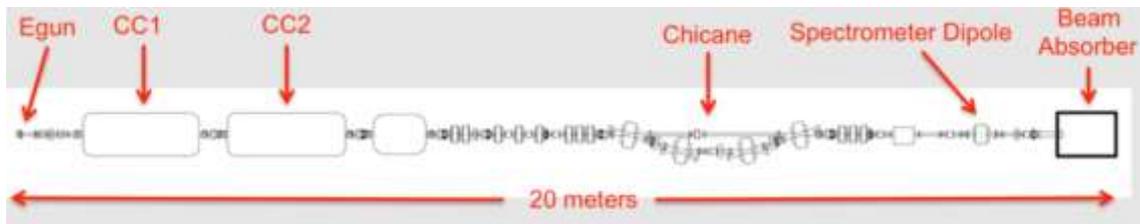

**Figure 11:** Overview of the photoinjector. The legend is L1, L2: solenoidal lenses, CAV1, CAV2 superconducting TESLA cavities, BC1: magnetic bunch compressor, DL: dogleg, green rectangles: quadrupole magnets, RFBT: round-to-flat-beam transformer.

The high energy section begins with a TESLA Type IV ILC-style cryomodule, which has been conditioned previously to 31.5 MeV/cavity at 2 K [12] to yield a total acceleration of 250 MeV and, therefore, provide up to 300 MeV electron beam. The cryomodule is driven by a single 1.3 GHz 5 MW klystron, and RF power is distributed to the eight separate cavities by variable tap-offs in the waveguide structure alongside the cryomodule (Fig. 13). The high energy transport consists of a FODO lattice with matching sections to direct the electrons through a dog-leg either to the high energy absorber or IOTA. A summary of the IOTA electron injector beam parameters is given in Table 4. Downstream of the linac is the test beam line section, which consists of an array of multiple high-energy beam lines that transport the electron beam from the accelerating cryomodules to one of two beam absorbers. Each absorber is capable of dissipating up to 75 kW of power and is surrounded by a large steel and concrete shielding dump.

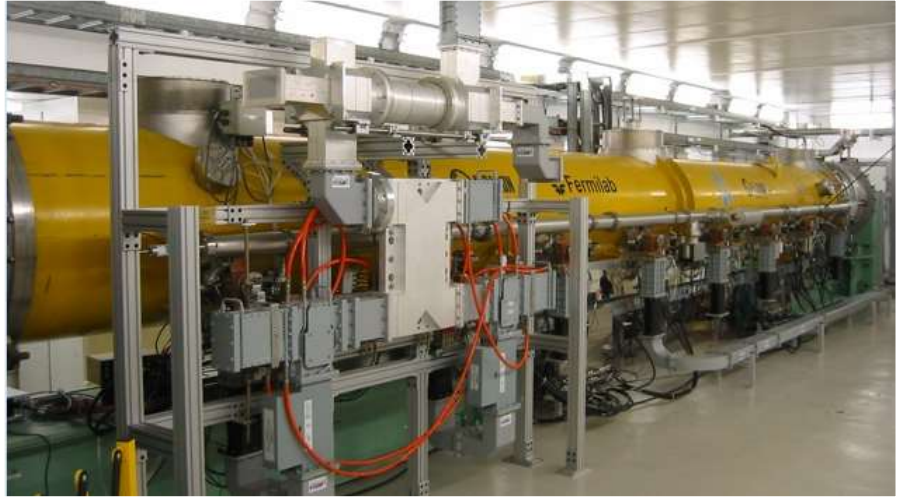

**Figure 13:** The first 1.3 GHz SRF cryomodule installed in the FAST/IOTA facility.

**Table 4:** Beam parameters of the IOTA electron injector.

| Parameter | Value |
|---|---|
| Beam Energy | 20 MeV – 300 MeV |
| Bunch Charge | < 10 fC – 3.2 nC per pulse |
| Bunch Train (Macropulse) | 0.5 – 9 MHz for up to 1 ms (3000 bunches, 3 MHz nominal) |
| Bunch Train Frequency | 1 – 5 Hz |
| Bunch Length | Range: 0.9 – 70 ps (Nominal: 5 ps) |
| Bunch Emittance 50 MeV, 50 pC/pulse | Horz: $1.6 \pm 0.2$ μm  Vert: $3.4 \pm 0.1$ μm |

## 2.3 Proton injector

As described above, an electron beam will initially be used to probe the optics of the IOTA ring. However, because the space charge effects on the electron beam will be negligible, and space-charge dominated proton beam will be used for most of the direct tests of the inherent stability. The IOTA proton injector will reuse the 70 MeV/c (2.5 MeV kinetic energy), low duty factor RFQ that was originally built and commissioned for Fermilab's High Intensity Neutrino Source (HINS) program [13]. This program was terminated when a leak in the water cooling system prevented this RFQ from reaching its design 1% duty factor. In our application, it is planned to use the RFQ for 1.7 μsec pulse, at 1 Hz maximum repetition rate, so no water cooling of the RFQ will be needed. The RFQ itself is a four vane design, operating at 325 MHz (Fig. 14). The source consists of a 50 kV filament proton source, capable of delivering 8 mA.

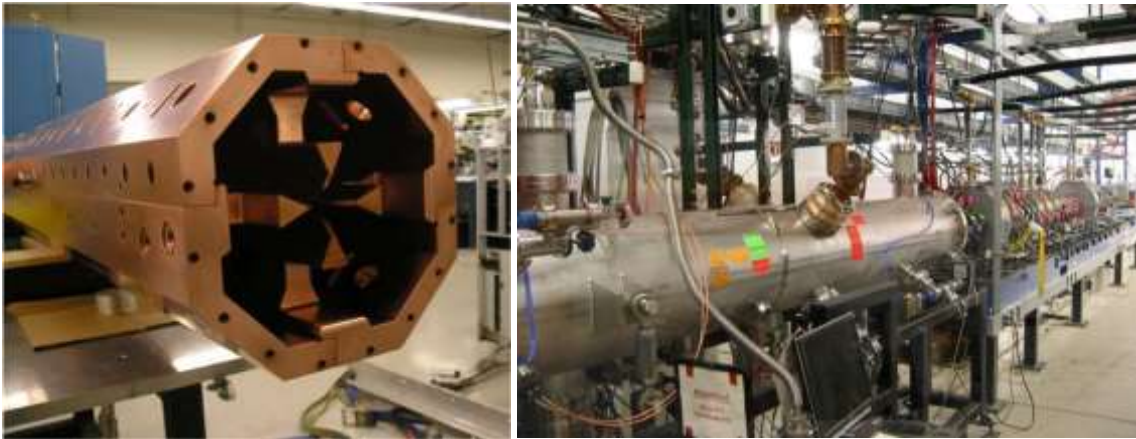

**Figure 14:** The bare four vane 2.5 MeV 325 MHz pulsed RFQ is shown at left, while at right is shown the RFQ in the vacuum vessel.

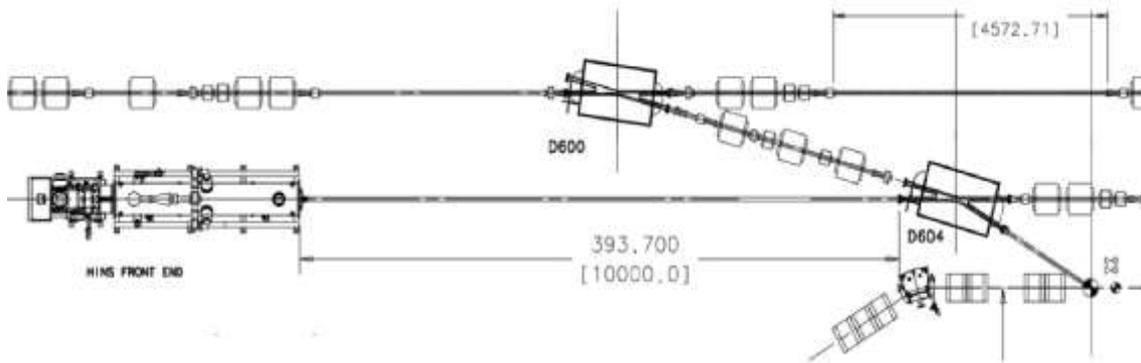

**Figure 15:** Configuration for proton injection into the IOTA ring. Because the direction is the same as that for electrons, all polarities will need to be reversed.

**Table 5:** Parameters of the IOTA proton RFQ injector.

| Parameter | Value |
|---|---|
| Beam energy, E | 2.5 MeV |
| Momentum, p | 68.5 MeV/c |
| dp/p | 0.1% (after debuncher) |
| Relativistic $\beta$ | 0.073 |
| Beam Rigidity | 0.23 T-m |
| RF frequency | 325 MHz |
| Beam Current | 8 mA |
| Pulse duration | 1.77 μsec (for one turn injection) |
| Total protons per pulse | $9 \times 10^{10}$ |
| RMS Emittance | 4 π-mm-mr (unnormalized) |
| Space-charge tune shift parameter | -.51 x (bunching factor) |
| Injection repetition rate | < 1 Hz |

Table 5 presents the key parameters of the IOTA ring operating with the 2.5 MeV RFQ and Fig.15 shows the proton injection configuration. The dipole that is used to switch electrons between the IOTA ring and the dump will also serve to switch protons into the ring. The 2.5 MeV energy of the protons corresponds to roughly half the nominal rigidity of the IOTA electron beam, but the velocity is only 0.073$c$, which presents some challenges. We plan to inject protons in the same direction as the electrons, using the same Lambertson magnet and kicker. Obviously, this means all field polarities will have to be reversed, and the pulse length for the kicker will need to be selectable. The RFQ will be followed by a 325 MHz debunching cavity to reduce the momentum spread somewhat. Nevertheless, because of the low $\beta$ of the proton beam, the momentum spread will cause the beam to completely debunch within the first turn or so, making the 325 MHz bunch structure irrelevant.

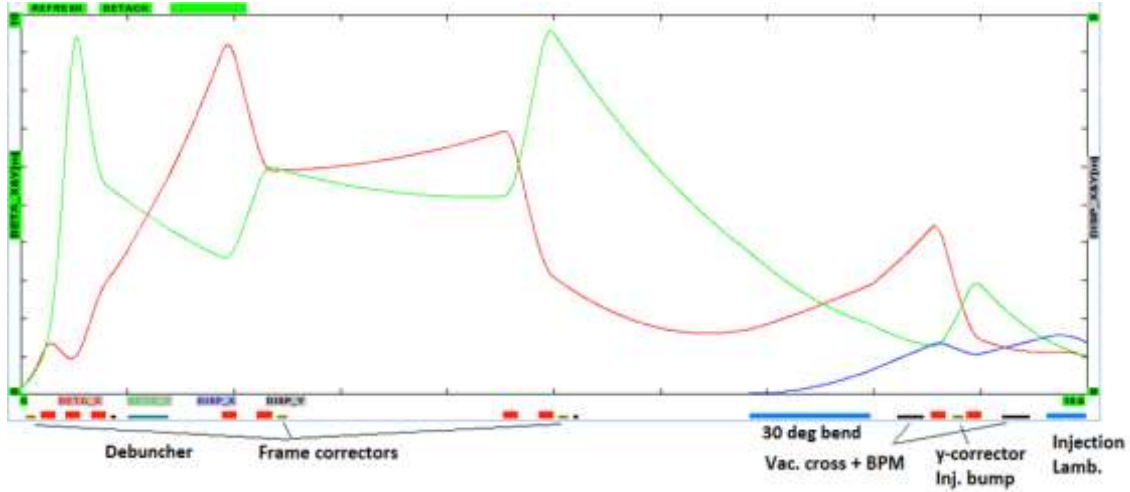

**Figure 16:** Optical design for the proton transfer line into the IOTA ring.

Fig. 16 shows the optics for the proton transfer line. The main design challenges were to accommodate the small aperture of the upstream debunching cavity and to match the lattice functions at injection. Unfortunately, because of the short distance between the selection dipole and the injection Lambertson magnet, it was not possible to make the resulting dogleg fully achromatic, meaning there is some residual dispersion at the injection point.

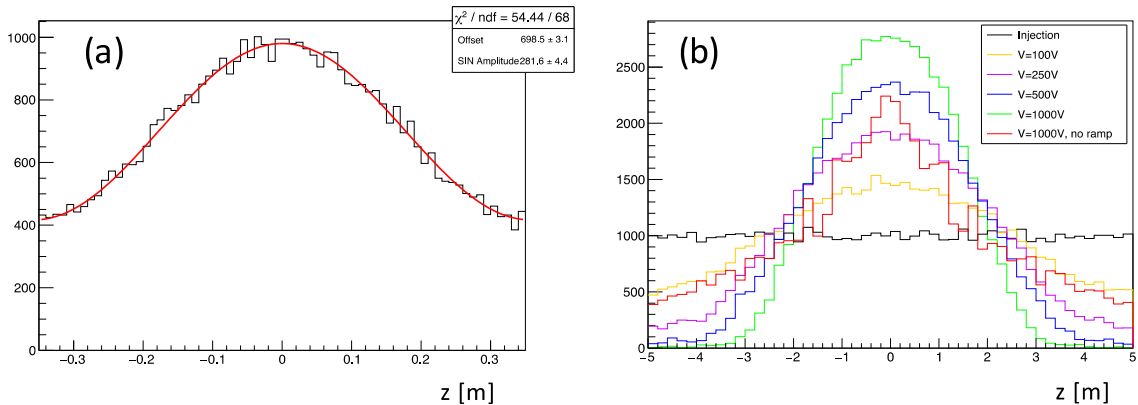

**Figure 17**: Proton bunching at (a) 30 MHz ($h$=56) and (b) 2.18 MHz ($h$=4).

The difference in the periods of the electrons and protons presents a challenge for both the RF system and for instrumentation. The RF system will initially operate at harmonic $h$=4, or 30 MHz, for electrons. The nearest harmonic for protons would be 56; however, as shown in Fig. 17(a), the planned voltage of the RF system is not nearly enough to fully bunch the proton beam at this frequency [14]. For this reason, the RF system has been designed to also operate at 2.18

MHz, which corresponds to $h=4$ for protons. As shown in Fig. 17 (b), this can fully bunch the proton beam at a reasonable voltage of 500V.

The different RF frequency presents something of a challenge for the BPMs, which will initially be designed to operate at 30 MHz. One solution would be to bunch the beam at 2.18 MHz, but then modulate it with the 30 MHz RF to generate a signal in the BPMs. Unfortunately, our simulations show that the signal from such a modulated beam would only correspond to about 10% of the total protons, which would significantly degrade the position resolution. Another solution is to modify the BPMs electronics to work with both frequencies. Among other things, this would require changing the input coupling to capacitive.

## 2.4 Beam instrumentation and controls

Given the research nature of the IOTA ring and its injectors, the accelerators will be equipped with an extensive suite of beam diagnostics systems needed to assure reliable operation and also to facilitate various experimental beam studies.

### 2.4.1 Beam Position Monitors

The beam position monitors (BPMs) in both the electron injector and the IOTA ring consist of 11-mm-diameter button electrodes with a subminiature type-A (SMA) feedthrough in a 47.5-mm inner-diameter housing. The BPM design is an upgrade of the system used for the ATF accelerator at KEK [15]. Each instance contains 4 buttons to provide both horizontal and vertical measurements. There are 20 BPMs upstream of the cryomodule and 20 downstream of it. The IOTA ring contains 20 standard BPMs, plus one with a wider aperture near the injection point. There is in-tunnel signal conditioning, and out-of-tunnel downmix electronics for 650 MHz, with digital signal processing via custom 12-channel, 250-MS/s, VME/VXS digitizers. The system can provide position resolutions of <50 μm for a single 3-nC bunch, and ~1 micron for 3000 bunches. It provides both relative intensity and phase and is sensitive down to ~50 pC/bunch.

### 2.4.2 Faraday Cup

Just downstream of the RF photoinjector's gun is a Faraday cup that can be inserted to measure beam current (Fig. 10). It is constructed from OFHC copper with dimensions of 1" x 1.75" x 0.5". The signal is directed out of the UHV system by means of a 0.051"-diameter stainless steel wire. The Faraday cup is inserted remotely via a horizontal pneumatic actuator.

### 2.4.3 Toroids and Wall Current Monitors

Beam intensity measurements are made along the electron injector using toroids, and high-bandwidth wall current monitors (WCM) developed for EMMA [16]. The WCMs feature a compact 50-mm-width design, including flanges, and have a 48-mm-diameter bore. The frequency response ranges from ~16 kHz, due to gap resistance and ferrite inductance, up to ~4 GHz limited by the microwave cutoff of high-order modes. They provide bunch-by-bunch intensity via an analog integrator and custom digitizer. Dual channels are used to sample the signal and background for each bunch. A 3-nC bunch charge results in a 1% accuracy in relative bunch intensity. The toroids are also a compact design and provide the absolute calibration for the WCMs. There are 2 WCMs upstream of the cryomodule and one planned for the IOTA ring. There are 2 toroids upstream of the cryomodule and 2 downstream.

### 2.4.4 Transverse Profile Monitors

The transverse profile monitors (TPM) in the electron injector were manufactured by Radiabeam. They consist of a 4-position holder containing 3 insertable screens: a YAG:Ce crystal oriented normal to the beam with a 45° mirror for light extraction, an Al-coated Si wafer for OTR generation (also with a 45° mirror), and a calibration target. The fourth position contains a cylindrical tube to reduce wakefields. The normal-incidence orientation of the YAG:Ce screen serves to reduce the depth of focus and crystal thickness effects on the resolution. Insertion of the 4-position holder is accomplished by a pair of pneumatic actuators and the light from the screens is transported and focused on a camera which provides the images of the beam for processing. There are 5 TPMs upstream of the cryomodule and 5 downstream. The optical transport contains, in sequence, a 150-mm achromatic lens, bending mirror, 150-mm achromatic lens, 200-mm achromatic lens, and a 75-mm camera lens. In addition, there are two remote-controlled filter wheels positioned before the camera lens containing four neutral density filters (ND0.3, ND1.0, ND2.0, ND3.0), two wavelength filters (400 ± 35 nm, 550 ± 35 nm), and a vertical and horizontal polarizer. The camera is a Prosilica GC2450, with a 5 mega-pixel imager and GigE-based readout. It has a 25-μs minimum shutter duration.

*2.4.5 Emittance Station*

Transverse beam emittances are determined with a slit-based measuring station that utilizes a 1" x 1" x 0.5 mm tungsten multi-slit mask with nine 40-micron-wide slits. The emittance station has both horizontal and vertical multi-slit masks on stepper motors as well as a YAG:Ce crystal located in the same holder as one of the slits so that the spot size of the electron beam can be measured. Downstream of the slits, a TPM will image the beam that made it through the slits on a YAG:Ce crystal to obtain the divergence of the beam. In addition to the multi-slit masks, a single 40 micron wide slit mask also resides in each holder to allow the potential disambiguation of a more complicated phase space.

*2.4.6 Beam Loss Monitors*

The beam loss monitor (BLM) system [17] is designed to provide both machine protection (alarm signal) and diagnostic functions for the machine, allowing tune-up and monitoring of beam operations while machine protection is integrating the same signal. The system incorporates linear, logarithmic and integrating amplifiers and has instantaneous read-back of beam losses. The loss monitors are based on photomultiplier tubes (PMTs) with a fast response of $\ll 1$ μs. The readout electronics are VME-based and can handle a macro-pulse repetition rate of up to 5 Hz. The sampling frequency of the 12-bit ADC is 3 MHz with a minimum buffer length of 1 ms. There are 10 BLMs upstream of the cryomodule and 25 downstream.

*2.4.7 Longitudinal Diagnostics*

Longitudinal diagnostics in the electron injector are provided by a streak camera, THz interferometer, and ceramic gap THz emission monitor. Optical and THz wavelength light can be collected from either of the last two dipoles in the bunch compressor (synchrotron radiation), or from a transition radiation screen located downstream of the bunch compressor [18].

*2.4.8 Streak Camera*

The streak camera is a Hamamatsu C5680 dual-sweep camera with a Prosilica GigE camera as the imager. The synchroscan vertical deflection unit in the mainframe of the streak camera is synchronized to 81.25 MHz from the master oscillator of the electron injector. The fastest range R1 provides about 0.6 to 1 ps temporal resolution (sigma). This unit provides low jitter (~1 ps) of streak images compared to a single-sweep unit that has 20-ps internal trigger jitter plus 100-ps trigger jitter from the delay unit. It also allows synchronous summing of signals from the micropulse train with about 200 fs phase stability within the train of micropulses. Gating of the MCP in the streak tube allows selection of a 250-ns window, corresponding to a single micropulse from the 3-MHz pulse train.

*2.4.9 Martin-Puplett Interferometer*

The Martin-Puplett interferometer is a THz polarizing interferometer with wire grids for the polarizers and splitter, commonly used for short bunch length measurements. Coherent THz radiation originating from one of the sources (synchrotron or transition), enters the interferometer and is split and recombined with a phase difference. Orthogonal linear polarizations of the resulting signal are analyzed by a pair of pyroelectric detectors. The intensity is accumulated for various phase differences, accomplished by moving a mirror in one leg of the split signal before recombining. This intensity distribution is the cosine transform of the frequency spectrum of the bunch. The sensitivity of the interferometer is expected to range from 0.2 ps to 1 ps rms bunch length. The lower cutoff is governed by the geometry of the wire grids, which consist of Tungsten wires with a diameter of 10 μm and a period of 45 μm, while the upper cutoff is determined by the low frequency characteristics of both the source and the detector.

*2.4.10 Ceramic Gap Monitor*

A ceramic-gap monitor is installed to measure the GHz/THz emission from an unshielded ceramic gap. A narrow band Schottky diode detector attached to a waveguide is used to measure the intensity in a given frequency band. This intensity can in principle be absolutely normalized if the bunch length can be made short enough to surpass the detector band.

*2.4.11 Allison Scanner*

An Allison-type emittance monitor [19] is used in the proton injector to measure transverse emittances. This device consists of a pair of sequential slits with a deflecting field between them. The beam impacts the entrance slit of the scanner and the transmitted beam passes between a pair of electrically charged deflecting plates. The surface of the plates are stair-cased to prevent impacting particles (and subsequent scattered secondaries) from passing through the exit slit. At specific plate voltages a portion of the transmitted beam is deflected such that it is transmitted through the exit slit and onto a collector. The collector current as a function of the plate voltage is proportional to the beam phase-space density as a function of $x'$. The scanner box is stepped through the beam in order to obtain a full two-dimensional map of the beam in $x-x'$ phase space.

*2.4.12 Synchrotron Radiation Diagnostics*

Profile measurements of electrons for the IOTA ring are provided by imaging of the synchrotron radiation emitted from the dipole magnets. The design of the beampipe dictates that the collected light originates from the entrance of the 30° magnets and from the center of the 60° magnets. The light is transported to a light-tight box positioned on top of the dipole containing a GigE-type CCD camera. The optical trajectory is controlled by a remotely-moveable turning mirror and an iris which controls the length of the emission source seen by the camera. Focusing is done via a 400 mm focal length lens resulting in a magnification of 0.837.

*2.4.13 Ionization Profile Monitor*

Another possible profile device for protons, either in the injector or the ring, is an ionization profile monitor (IPM), of which two are available from the decommissioned Tevatron collider ring [20]. This device collects either the ions or the electrons from ionzations induced by the beam and produces a beam profile from them. The amount of signal is proportional to the ionization cross section and the gas density. The gas source can either be the residual gas in the beampipe, or a purposely injected gas. In IOTA, the residual gas pressure will be very low leading potentially to a very small signal. Fortunately, since the energy of the protons is very low, the cross section is much higher, and roughly cancels the effect of the low gas pressure.

## 3. Beam Physics Experiments at IOTA

### 3.1 Integrable optics with nonlinear magnets

Lattice design of all present accelerators incorporates dipole magnets to bend particle trajectory and quadrupoles to keep particles stable around the reference orbit. These are "linear" elements because the transverse force is proportional to the particle displacement, $x$ and $y$. This linearity results (after the action-phase variable transformation) in a Hamiltonian of the following type:

$$H(J_1, J_2) = \nu_x J_1 + \nu_y J_2, \quad (1)$$

where $\nu_x$ and $\nu_y$ are betatron tunes and $J_1$ and $J_2$ are actions. This is an integrable Hamiltonian. The drawback of this Hamiltonian is that the betatron tunes are constant for all particles regardless of their action values. It has been known since early 1960-s that the spread of betatron tunes is extremely beneficial for beam stability due to the so-called Landau damping. However, because the Hamiltonian (1) is linear, any attempt to add non-linear elements (sextupoles, octupoles) to the accelerator generally results in a reduction of its dynamic aperture, resonant behavior and particle loss. A breakthrough in understanding of stability of Hamiltonian systems, close to integrable, was made by Nekhoroshev [21]. He considered a perturbed Hamiltonian system:

$$H = h(J_1, J_2) + \varepsilon q(J_1, J_2, q_1, q_2), \quad (2)$$

where $h$ and $q$ are analytic functions and $\varepsilon$ is a small perturbation parameter. It was proved that under certain conditions on the function $h$, the perturbed system (2) remains stable for an exponentially long time. Functions $h$ satisfying such conditions are called steep functions with quasi-convex and convex being the steepest. In general, the determination of steepness is quite complex. One example of a non-steep function is a linear Hamiltonian Eq. (1).

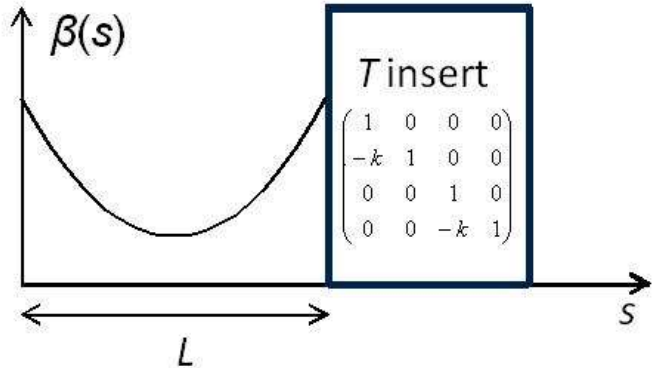

**Figure 18**: An element of periodicity: a drift space with equal beta-functions followed by a T-insert.

In Ref. [22] three examples of nonlinear accelerator lattices were proposed. One of the lattices, which results in a steep (convex) Hamiltonian is being implemented in IOTA as follows. Consider an element of lattice periodicity consisting of two parts: (1) a drift space, $L$, with exactly equal horizontal and vertical beta-functions $\beta$, followed by (2) an optics insert, $T$, which is comprised of linear elements and has the transfer matrix of a thin axially symmetric lens (Figure 18).

It was shown in [22] that it is possible to construct a nonlinear dynamical system that has two integrals of the motion by introducing an additional transverse magnetic field along the drift space $L$. The potential, $V(x, y, s)$, associated with this field satisfies the Laplace equation, $\Delta V = 0$ and hence can be implemented with a conventional electromagnet.

The Hamiltonian for a particle moving in the drift space $L$ with an additional potential $V$ is:

$$H = \frac{p_x^2 + p_y^2}{2} + \frac{x^2 + y^2}{2} + \beta(\psi) V\left(x\sqrt{\beta(\psi)}, y\sqrt{\beta(\psi)}, s(\psi)\right), \quad (3)$$

Where $\psi$ is the "new time" variable defined as the betatron phase $\psi' = 1/\beta(s)$.

The potential in equation (3) can be chosen such that it is time-independent. This results in a time-independent Hamiltonian (3) and hence the first integral of the motion. The second integral of the motion is a quadratic function of the momenta [22]:

$$I = (xp_y - yp_x)^2 + c^2 p_x^2 + \frac{2c^2 t \cdot \xi \eta}{\xi^2 - \eta^2}\left(\eta\sqrt{\xi^2 - 1}\cosh^{-1}(\xi) + \xi\sqrt{\eta^2 - 1}\left(\frac{\pi}{2} + \cosh^{-1}(\eta)\right)\right), \quad (4)$$

where

$$\xi = \frac{\sqrt{(x+c)^2 + y^2} + \sqrt{(x-c)^2 + y^2}}{2c}, \eta = \frac{\sqrt{(x+c)^2 + y^2} - \sqrt{(x-c)^2 + y^2}}{2c} \quad (5)$$

are the elliptic variables and $t$, $c$ are arbitrary constants.

The potential $V$ is presented in the following way:

$$\beta(\psi) V\left(x\sqrt{\beta(\psi)}, y\sqrt{\beta(\psi)}, s(\psi)\right) = \frac{x^2 + y^2}{2} + t \frac{\xi\sqrt{\xi^2 - 1}\cosh^{-1}(\xi) + \eta\sqrt{\eta^2 - 1}\left(\frac{\pi}{2} + \cosh^{-1}(\eta)\right)}{\xi^2 - \eta^2}. \quad (6)$$

Figure 19 presents a contour plot of the potential energy Eq. (6) for $c = 1$ and $t = 0.4$.

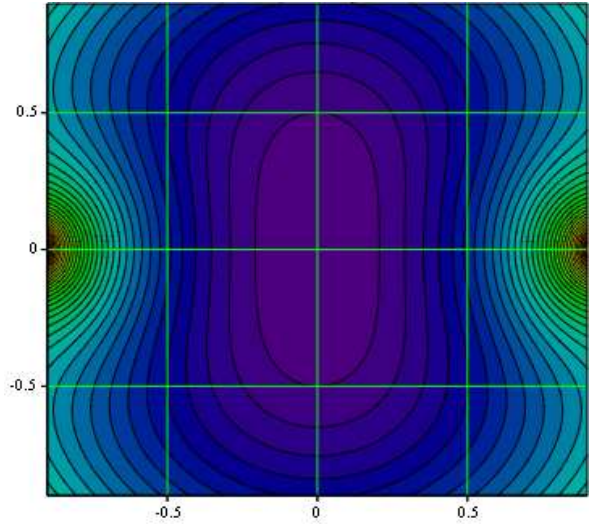

**Figure 19:** Contour plot of the potential energy Eq. (10) with $c = 1$ and $t = 0.4$. The repulsive singularities are located at $x = \pm c$ and $y = 0$.

The multipole expansion of this potential for $c = 1$ is as follows:

$$U(x, y) \approx \frac{x^2 + y^2}{2} + t\,\mathfrak{Re}\left((x + iy)^2 + \frac{2}{3}(x + iy)^4 + \frac{8}{15}(x + iy)^6 + \frac{16}{35}(x + iy)^8 + \cdots\right), \quad (7)$$

where *t* is the magnitude of the nonlinear potential.

The betatron phase advance over the drift space *L*, is limited to 0.5 (in units of $2\pi$). The phase advance in *T*-insert must be a multiple of 0.5. This makes the maximum full tune of one element of periodicity 0.5+0.5×*n* (*n* being an integer). The theoretical maximum attainable nonlinear tune shift per cell is ~0.5 for mode 1 and ~0.25 for mode 2. Expressed in terms of the full betatron tune per cell, this tune shift can reach 50% (0.5/(0.5+0.5)).

Numerical simulations with single and multi particle tracking codes were carried out in order to determine the tune spread that can be achieved in a machine built according to the above recipe [24,25,26]. Various imperfections were taken into account, such as the perturbations of T-insert lattice, synchrotron oscillations, and other machine nonlinearities. In Figure 20 a result of one of the simulations is presented. The tune footprint obtained with Frequency Map Analysis [27] demonstrates that vertical tune spread exceeding 1 can be achieved with four elements of periodicity and very little resonances are caused by imperfections.

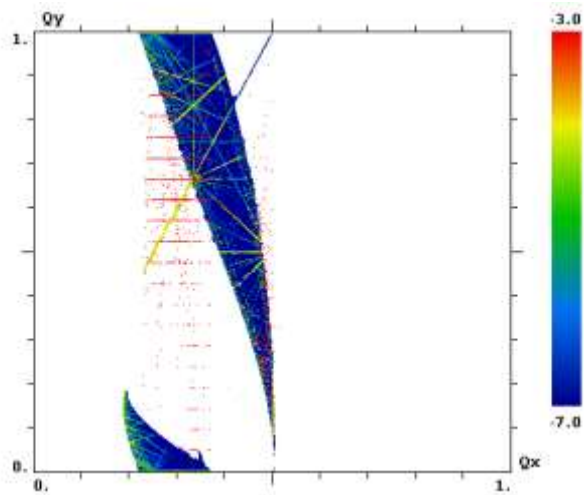

**Figure 20:** Beam tune footprint for $\nu_0 = 0.3$, four elements of periodicity, $t = 0.4$. Simulation with Lifetrac Frequency Map Analysis

The condition of the Hamiltonian time-independence requires that the nonlinear potential must *continuously* change along the length of the nonlinear section (Eq. 3). The potential is defined by two parameters: the strength parameter *t*, and the geometric parameter *c*, which represents the distance between the singularities, or the element aperture (Eq. 6). The geometric parameter *c* scales as the square root of $\beta$-function in the nonlinear straight section, and the strength parameter *t* scales as $1/\beta$. Since it is not practical to manufacture a magnet with complex varying aperture, the adopted approach approximates the continuously varying potential with 18 "thin" magnets of constant aperture (Fig. 21).

The magnetic potential can be expanded into multipole series Eq. (7). However, this expansion is only valid inside the $r = \sqrt{x^2 + y^2} < c$ circle. Vertical oscillation amplitudes $y > c$ are essential for achievement of large tune spread. By proper shaping of the magnetic poles it was possible to achieve good field quality in the region $x < c$, $y < 2 \times c$.

The IOTA nonlinear insert has the length of 1.8 m, the design point betatron phase advance in the drift is 0.3, which corresponds to the minimum beta-function of 0.6 m and the maximum beta-function of 2 m. Thus, the geometric parameter *c* of the nonlinear magnet varies from 8 mm to 14 mm, and the horizontal beam pipe aperture is between 12 mm and 21 mm (Fig. 21).

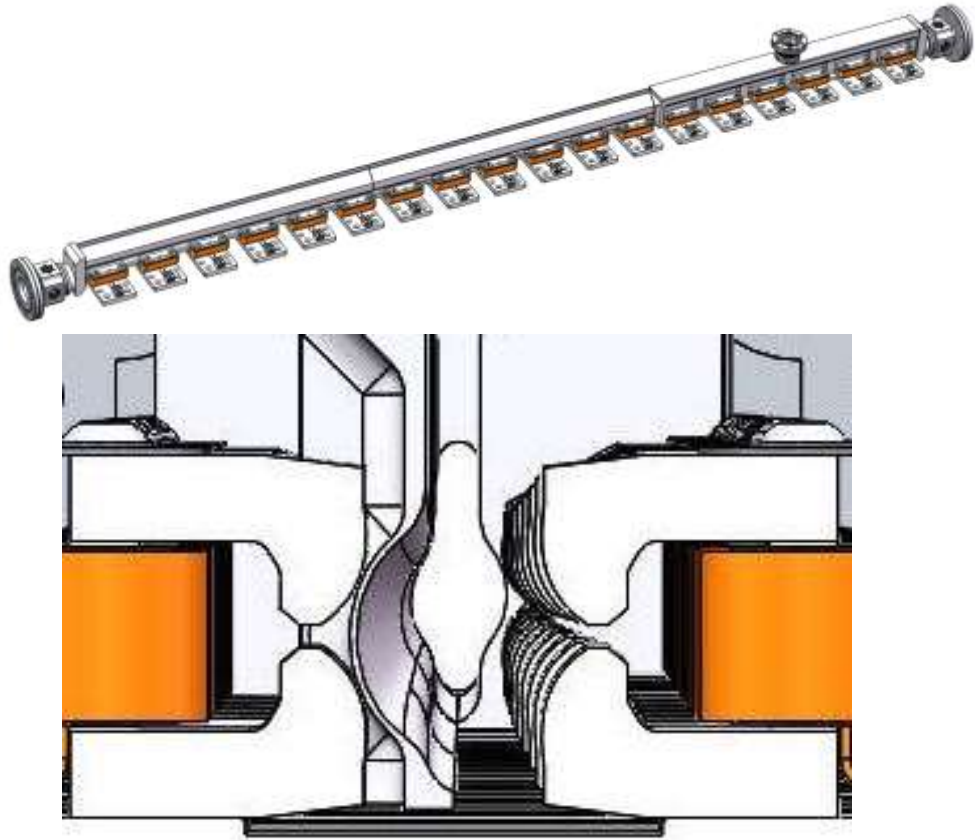

**Figure 21**. Solid model rendering of the 18-section IOTA nonlinear magnet (top). Magnet cross-section showing the vacuum chamber shape and magnetic poles (bottom). Images courtesy of RadiaBeam Technologies.

The goal of IOTA experiments on nonlinear integrable optics is to experimentally test the mathematical predictions. The unique opportunity to directly compare measurements with mathematical predictions for strongly nonlinear particle motion could fundamentally advance understanding of ring dynamics. Initial experiments will be conducted with low-charge electron bunches to emulate single particle dynamics with negligible space charge and other collective effects.

The experiments with electron beams will be used to demonstrate the achievement of large ($\approx 0.25$ for one element of periodicity) nonlinear amplitude-dependent tune shift without the degradation of dynamical aperture. The experiment will be carried out by "painting" the accelerator aperture with a "pencil" beam by kicking the circulating bunch in the transverse plane with the use of stripline kickers and then recording the turn-by-turn position of the bunch to a) measure the effect of the betatron tune variation with amplitude and b) demonstrate the conservation of the integrals of motion. The precision of the IOTA turn-by-turn BPM system (section 2.4) allows to resolve the conservation of the integrals down to a 1-percent level, which will allow to evaluate the conservation of 2D integrability. The experiments with electron beams provide a convenient environment because of the small emittance of the electron beam (the transverse beam size in the nonlinear insert is approximately 0.25 mm), and the relatively long synchrotron radiation damping time ($10^7$ turns). The experiment seeks to investigate the stability of nonlinear system to perturbations: chromatic effects, effect of synchrotron oscillations, lattice distortions.

For the planned proton beam experiments, a single bunch train will be injected from an RFQ, with the micro-bunches allowed to debunch longitudinally until they fill the ring uniformly. Simulations to date have used the Synergia framework [28, 29], ignoring RF effects, including 2D space charge forces that neglect any longitudinal effects. Fig. 22 shows the extent to which the Danilov and Nagaitsev invariants are conserved over 1 million turns in the IOTA lattice with a single nonlinear insert, for strength parameter $t$=0.4 and a 30 deg phase advance across the insert. These single-particle simulations assume zero momentum spread. The bounded, periodic nature of the fluctuations indicate that there are two nearby invariants of the motion and, hence, the dynamics remain integrable [30].

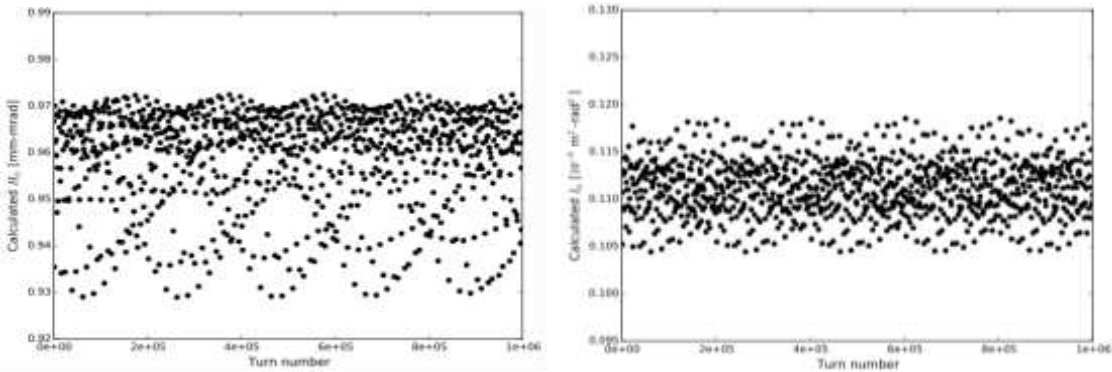

**Figure 22:** The 1st Danilov and Nagaitsev invariant (i.e. the Hamiltonian) is shown on the left, with the 2nd invariant shown on the right, calculated in the center of the nonlinear insert, showing 1.1% and 2.6% deviation from invariance, respectively, over a million turns. Synergia code was used for the simulations (Plots provided courtesy of N. Cook and S.D. Webb).

The small deviations from invariance seen in Fig. 22 have been explained via Lie perturbation theory (unpublished, the calculation is similar to [31]). The rms deviations are predicted to increase quadratically with the phase advance across the nonlinear insert and approximately linearly with emittance – showing quantitative agreement with Synergia simulations that linearize the IOTA lattice. Experimentally, it should be possible to observe the loss of integrability as the emittance is increased (e.g. through variable scraping of the RFQ pulses during injection), or as the phase advance across the nonlinear insert is increased (requires unique quad settings for each choice of phase advance).

**3.2 Integrable optics with nonlinear electron lenses**

In an electron lens, the electromagnetic field generated by a pulsed, magnetically confined, low-energy electron beam is used to actively manipulate the dynamics of the circulating beam [32, 33]. Electron lenses have a wide range of applications in beam physics [34, 35, 36, 37, 38, 39, 40, 41, 42, 43, 44, 45]. In particular, they can be used as nonlinear elements with a tunable transverse kick dependence as a function of betatron amplitude [34, 46, 47, 48].

The layout of the IOTA electron lens is shown in Figure 23. The electron beam, generated by a thermionic cathode, is confined and transported by strong axial magnetic fields through the overlap region, where it interacts with the circulating beam. It is then steered into a collector.

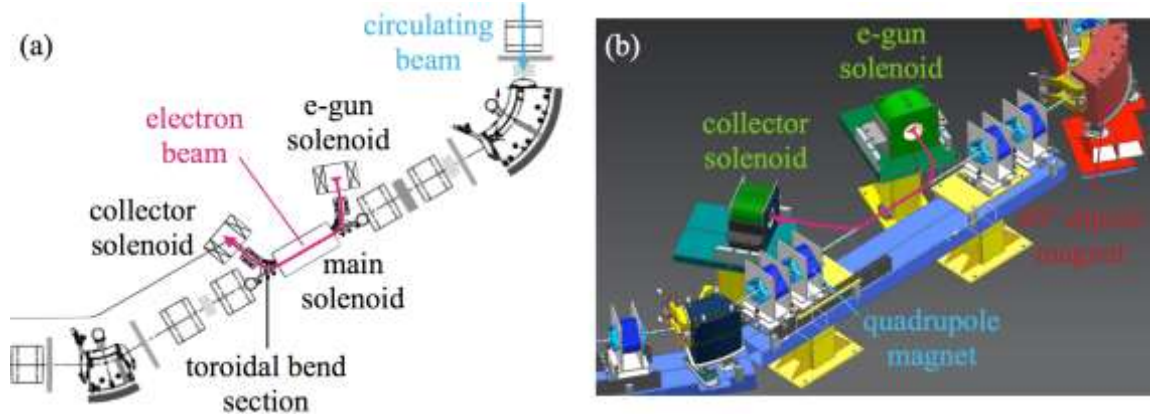

**Figure 23**. Layout of the electron lens in the IOTA ring. (a) Schematic plan view. (b) Solid drawing of selected components.

Let us consider the main quantities characterizing an electron lens. The cathode-anode voltage $V$ determines the velocity $v_e = \beta_e c$ of the electrons in the device, which is assumed to have length $L$ and to be located in a region of the ring with lattice amplitude function $\beta_{x,y}$. When acting on a circulating beam with magnetic rigidity $(B\rho)$ and velocity $v_z = \beta_z c$, the linear focusing strength $k_e$ for circulating particles with small betatron amplitudes is proportional to the electron current density on axis $j_0$:

$$k_e = 2\pi \frac{j_0 L (1\pm\beta_e\beta_z)}{(B\rho) \beta_e \beta_z c^2}. \quad (16)$$

The "+" sign applies when the beams are counter-propagating, and the electric and magnetic forces act in the same direction. For small strengths and away from the half-integer resonances, these kicks translate into the tune shift

$$\Delta\nu = \frac{\beta_{x,y} j_0 L (1\pm\beta_e\beta_z)}{2 (B\rho) \beta_e \beta_z c^2} \quad (17)$$

for particles circulating near the axis.

There are two concepts of electron lenses for nonlinear integrable optics: thin radial kick of McMillan type; and thick axially-symmetric nonlinear lens in constant amplitude function.

*3.2.1 Thin Radial Kick of McMillan Type*

The integrability of axially symmetric thin-lens kicks was studied in 1 dimension by McMillan [49, 50]. It was then extended to 2 dimensions [51] and experimentally applied to the beam-beam force, resulting in an improvement of the performance of colliders [52]. Let us consider an electron lens with a specific radial current-density distribution:

$$j(r) = j_0 \frac{a^4}{(r^2+a^2)^2}, \quad (18)$$

where $j_0$ is the current density on axis and $a$ is a constant parameter (the effective radius). The total current is $I_e = j_0 \pi a^2$. The circulating beam experiences nonlinear transverse kicks as a function of amplitude:

$$\theta(r) = k_e r \frac{a^2}{r^2+a^2}. \quad (19)$$

For such a radial dependence of the kick, if the element is thin ($L \ll \beta_{x,y}$) and if the betatron phase advance in the rest of the ring is near an odd multiple of $\pi/2$, then there are 2 independent invariants of motion in the 4-dimensional transverse phase space. Neglecting longitudinal effects, all particle trajectories are regular and bounded. The achievable nonlinear tune spread $\Delta\nu$ (i.e., the tune difference between small and large amplitude particles) is of the order of $(\beta_{x,y} k_e/4\pi)$. A more general expression applies when taking into account machine coupling and the electron-lens solenoid. For the thin McMillan lens, it is critical to achieve and preserve the desired current-density profile.

*3.2.2 Axially Symmetric Kick in Constant Beta Function*

The concept of axially symmetric thick-lens kicks relies on a section of the ring with constant and equal amplitude functions. This can be achieved with a solenoid with axial field $B_z = 2(B\rho)/\beta_{lat}$ to provide focusing for the circulating beam and lattice functions $\beta_{lat} = \beta_x = \beta_y$. The same solenoid magnetically confines the low-energy beam in the electron lens. In this case, any axially symmetric electron-lens current distribution $j(r)$ generates two conserved quantities, the Hamiltonian and the longitudinal component of the angular momentum, as long as the betatron phase advance in the rest of the ring is an integer multiple of $\pi$. At large electron beam currents in the electron lens, the focusing of the electron beam itself dominates over the solenoid focusing and can be the source of the constant amplitude functions. Because the machine operates near the integer or half integer resonances, the achievable tune spread in this case is of the order of $L/(2\pi\beta_{lat})$. This scenario benefits from long solenoids and low beta functions, and it is insensitive to the shape of the current-density distribution in the electron lens. Although in IOTA the achievable tune spread is smaller in this case than it is in the McMillan case, this scenario is more robust and will probably be the first one to be studied experimentally, using existing Gaussian or similar electron guns.

*3.2.3 Experimental Design and Apparatus*

For demonstrating the nonlinear integrable optics concept with electron lenses in a real machine, there are several design considerations to take into account. The size of the electron beam should be compatible with the achievable resolution of the apparatus. Amplitude detuning and dynamic aperture of the ring will be measured by observing the turn-by-turn position and intensity of a circulating pencil beam with an equilibrium emittance $\varepsilon = 0.1$ μm (rms, geometrical) and size $\sigma_e = \sqrt{\beta_{lat}\varepsilon}$ at the electron lens. This size should be larger than the expected resolution of the beam position monitors, $\sigma_{BPM} \leq 0.1$ mm. For typical IOTA electron-lens lattices, $\beta_{lat} = 3$ m and $\sigma_e = 0.55$ mm, which satisfies this requirement. Moreover, it follows that the required axial field is $B_z = 2(B\rho)/\beta_{lat} = 0.33$ T, so that the solenoid can provide the required focusing at low electron beam currents. The aperture of the ring $A_{ring} = 24$ mm must be sufficient to contain a wide range of betatron amplitudes and detunings. Aperture and magnet field quality suggest a maximum tolerable orbit excursion of about $A_{max} = A_{ring}/2 = 12$ mm. Particles at small amplitudes will exhibit the maximum detuning $\Delta\nu$. The maximum excursion $A_{max}$ must be sufficient to accommodate particles with large amplitudes and small detunings compared to $\Delta\nu$. For the McMillan kick distribution, for instance, this can be achieved by requiring $a \leq A_{max}/6 = 2$ mm. For a typical electron lens with resistive solenoids, with $B_z = 0.33$ T in the main solenoid, one can operate at $B_g = 0.1$ T in the gun solenoid. Because of magnetic compression, this translates into a current-density distribution with $a_g = a\sqrt{B_z/B_g} = 3.6$ mm at the cathode. This parameter serves as an input to the design of the electron-gun assembly.

The achievable tune spread should be large enough to clearly demonstrate the effect of nonlinearities and detuning on beam lifetime and dynamical aperture. As a goal for the IOTA project, it was decided to set $\Delta\nu \geq 0.25$. This requirement imposes a constraint on the current density in the electron lens. For instance, with typical electron-lens parameters, $L = 1$ m, $\beta_e = 0.14$ (corresponding to a kinetic energy of 5 keV) and co-propagating beams[1], one obtains $j_0 = 14$ A/cm$^2$ and, for the McMillan distribution, a total current $I_e = 1.7$ A. These parameters are within the current state of the art. Typical parameters for the IOTA electron lens are shown in Table 7.

Table 7: Typical electron-lens parameters for IOTA.

| Parameter | Value |
|---|---|
| Cathode-anode voltage | 0.1 – 10 kV |
| Electron beam current | 5 mA – 5 A |
| Current density on axis | 0.1 – 12 A/cm$^2$ |
| Main solenoid length | 0.7 m |
| Main solenoid field | 0.1 – 0.8 T |
| Gun/collector solenoid fields | 0.1 – 0.4 T |
| Max. cathode radius | 15 mm |
| Lattice amplitude function | 0.5 – 10 m |
| Circulating beam size (rms) | 0.1 – 0.5 mm ($e^-$) |
|  | 1 – 5 mm ($p$) |

It may be challenging to transport these currents through a resistive electron lens while preserving the desired quality of the current-density profile. The effects of imperfections and of longitudinal fields must be investigated with numerical simulations and with experimental studies in the Fermilab electron-lens test stand. A detailed numerical study of the magnetic field configuration, single-particle trajectories, and electron beam dynamics with self-fields was presented in [53].

The electron-lens project builds upon the many years of experience in the construction and operation of electron lenses at Fermilab, and it can rely on several components that are already available at the laboratory, such as electron gun assemblies, resistive solenoids, collectors, high-voltage modulators, and power supplies of the two decommissioned Tevatron electron lenses. The relatively large instantaneous beam power will require provisions for pulsed operation of the electron lens compatible with the time structure of the IOTA circulating beam, as well as a depressed collector. The toroidal bends and the main solenoid need to be redesigned because of cost and infrastructure (resistive solenoid in IOTA vs. superconducting solenoid in Tevatron) and because of the tight spaces in the smaller ring. An overview of the apparatus was given in [54].

The total beam current will be measured at the collector. A diagnostic station with retractable devices will be installed upstream of the collector to measure the current-density profile. If possible, an additional station will be installed just downstream of the electron gun. The beam tube inside the main solenoid will be instrumented with segmented electrodes, to be used as electromagnetic pickups, ion-clearing plates and confinement electrodes. A microwave antenna to detect cyclotron frequencies is also foreseen to calibrate the magnetic fields and to estimate the electron beam density and temperature. For proton operations in IOTA, an electron-proton recombination detector based on microchannel plates will be installed after the first dipole downstream of the device. Some of these diagnostics tools are dictated by the novel combined use of the device as a nonlinear lens, electron cooler, and electron trap.

---

[1] We consider the co-propagating case because it is more conservative and because it will be necessary for electron cooling of protons (which will be injected in IOTA in the same direction as circulating electrons).

### 3.3 Space-charge compensation with electron lenses

Space-charge forces often cause emittance growth, instabilities, and beam loss. Mitigation schemes include solenoidal fields, halo scraping, or compensation with a plasma column of opposite charge [55, 56, 33]. Solenoid-based compensation in RF photoinjectors is widely used [56]. Although plasma-based space-charge compensation is commonly used in linacs, its implementation in rings has not been achieved yet. Charge neutralization over the circumference of the ring can be impractical. Moreover, local compensation schemes require high charge densities, which in turn can cause beam scattring, distortions of lattice focusing, and beam-plasma instabilities. In circular machines, partial neutralization was attempted, showing considerable improvements in proton intensity, but also strong proton-electron instabilities [57].

Because an electron lens is based upon magnetically confined electron beams, some of these effects can be mitigated. There are two ways an electron lens can be used as a space-charge compensator. One relies on an electron gun that generates the required charge distribution in transverse space and possibly in time, to reproduce the bunch shape of the circulating beam [58]. In the other scheme, the so-called 'electron column', the electrons are generated by ionization of the residual gas and trapped axially by electrodes and transversely by the solenoidal field, in a configuration similar to a Penning-Malmberg trap [59] (see also Section 3.4 below). In this latter case, the electron gun and collector are not necessary.

The physics of the interaction between circulating bunches and electron plasma is still a very open field of research [60]. However, the required gun, solenoid, and electrode parameters are similar to those of an electron lens, and therefore theory and experiment can be studied in IOTA.

### 3.4 Space-charge compensation with electron columns

The IOTA ring will be also used to study space charge compensation using so called "electron column" [61]. The compensation is achieved through trapping and controlling of intense negative charge of electrons, generated from residual gas ionizations by intense and stable circulating proton beams inside a strong solenoidal magnetic field (Fig. 24). The technique is similar to the electron lens SCC described above, except it does not require an external electron source and sophisticated transport system - the circulating proton beam itself is used to generate electrons and match them to the proton beam profile. The experimental test of the electron column SCC technique will be performed with high brightness, space-charge dominated proton beam in IOTA. The IOTA electron lens main solenoid and electrostatic electrodes inside it will be used to control accumulation of the ionization electrons.

Optimal compensation conditions for the electron column method could be obtained if the distributions of electrons and circulating beam are matched transversely, and, ideally, longitudinally. A strong solenoidal fields is supposed to assure the transverse matching of the trapped electrons while the trapping voltage on the electrostatic electrodes allows to control the total electron charge accumulated. The strong external magnetic field also prevents coherent *e-p* instabilities as it stabilizes the transverse motion of electrons in the column. At the same time, the magnetic field should not be too strong to allow ions to escape easily from the column, and not interfere with the processes of charge compensation. It also expected that the neutralization time can be fast enough that the accumulation can track the changes in the proton beam size due to, e.g., acceleration.

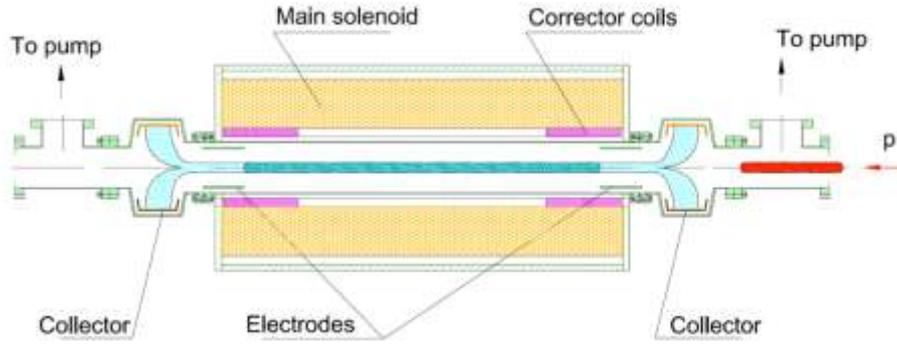

**Figure 24:** Schematic layout of the "electron column" experimental setup.

The WARP code [62] simulations of physical processes inside the electron column are used to understand the precise parameter ranges for the solenoidal magnetic field, electrodes' voltages, vacuum pressure and other experimental test parameters. The dominant charge generating processes in the electron column are the primary beam ionization $p + H_2 \rightarrow p + H_2^+ + e$ and the secondary ionization $e + H_2 \rightarrow H_2^+ + 2e$ [63]. For the IOTA proton beam parameters of 8 mA current and 70 MeV/c momentum, the electrostatic potential at the proton beam center is given by:

$$\phi = \frac{30I}{\beta} \sim 3.5 \text{ V}$$

(here $I$ is a beam current and $\beta$ is a relativistic velocity of protons). With ~1 m length of an electron column to compensate the proton space charge effect accumulated over ~40 m circumference of the IOTA ring, one needs the voltages on the electrodes of about 3.5 V × (40/1)= 140 V for a full space charge compensation. Initial study has been performed for a straight 1 m channel only, i.e., no proton circulation in the ring is considered [64]. The main knobs to control electron density profiles are solenoidal magnetic field (0 to 0.1 T), voltages on the left and right electrodes (0 to -200 V), and vacuum pressure ($10^{-3}$ to $10^{-5}$ Torr). The simulation time step is set to 15 psec which is less than the cyclotron period. In each time step, 100 macro particles of the proton beam – each equivalent to 7,500 actual particles - are injected.

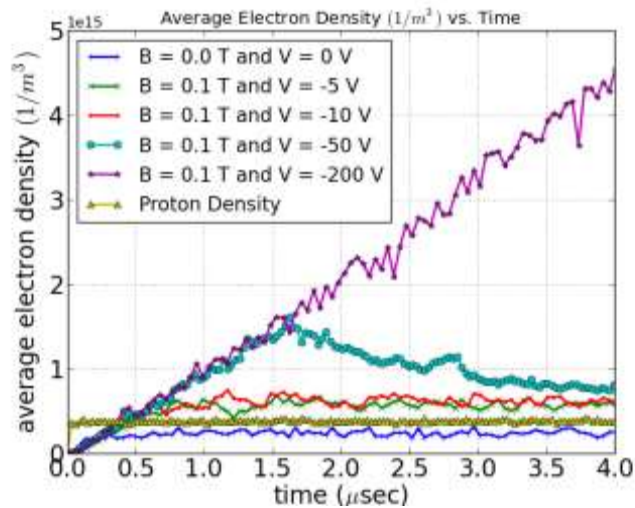

**Figure 25**: Neutralization time for various parameters.

Fig. 25 shows the comparison of the electron density variations in time for different control parameters as well as the proton density. If solenoid field and trapping voltages are turned on, electron densities grow in time until the charge neutralization is achieved. When voltages on electrodes are higher, the neutralization time gets longer, though in all scenarios is comparable to the revolution time in the IOTA ring ~1.8 μsec.

Without external magnetic field and trapping voltages, simulations result in the longitudinal densities of electrons lower than that of protons, i.e., under-compensation. As shown in Fig. 26, the transverse density profiles of the electrons are closely matched to those of the protons for $B_{sol} = 0.1$ T and $V_{electrode} = -5.0$ V. For higher voltages on electrodes, more electrons tend to be trapped between two electrodes inside the column; therefore, densities of electrons exceed those of protons. There are significant over-compensations for $B_{sol} = 0.1$ T and $V_{electrode} = -200.0$ V. When applied voltages are higher, the densities of $H_2^+$ are also noticeably increased, so it could affect the physical processes between protons and electrons.

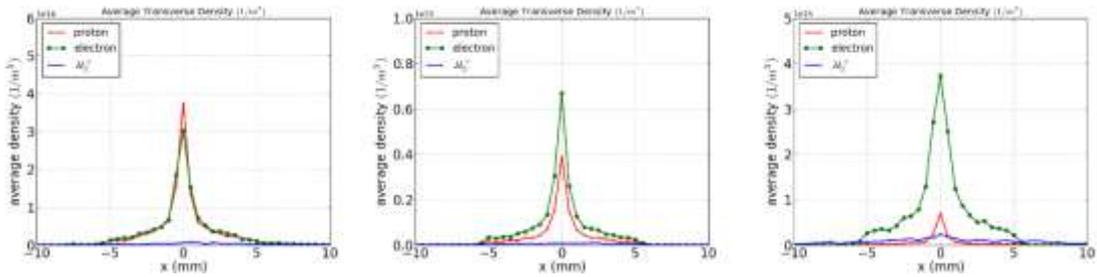

**Figure 26:** Transverse density profiles of proton beam, electrons and $H_2^+$ ions in the electron column under various $B_{sol}$ and $V_{electrode}$ : left – (0T, 0V); middle – (0.1T, -5V); right – (0.1T, -200V). The vacuum pressure is $10^{-3}$ Torr.

Simulation studies show that the density profile of the e-column can be tuned with axial *B*-field, electrode voltages, and vacuum pressure for partial/full/over compensations of space charge effects. Practical issues, such as ring dynamics of the primary proton beam with external focusing, and multiple passes through the electron column in the IOTA ring, will be investigated separately.

### 3.5 Optical stochastic cooling

Due to large longitudinal particle density, $N/\sigma_s$, the beam cooling times of few hours required for luminosity control in high energy hadron colliders cannot be achieved with traditional stochastic cooling technique in the microwave frequency range (~$10^9$-$10^{10}$ Hz). A practical scheme operating in the optical frequency range – optical stochastic cooling (OSC) - suggested in [65] allows the increase of cooling bandwidth by a few orders of magnitude to about ~$10^{14}$ Hz. In OSC, a particle emits electro-magnetic radiation in the first (pickup) wiggler. Then, the radiation amplified in an optical amplifier (OA) makes a longitudinal kick to the same particle in the second (kicker) wiggler, as shown in Fig. 27. A magnetic chicane is used to make space for the OA and to delay a particle so that to compensate for a delay of its radiation in the OA resulting in simultaneous arrival of the particle and its amplified radiation to the kicker wiggler. A particle passage through the chicane has a coordinate-dependent correction of particle longitudinal position which, consequently, results in a correction of relative particle momentum, $\delta p/p$, with amplitude $\xi_0$ so that:

$$\delta p / p = -\xi_0 \sin(k \Delta s) \ . \qquad (20)$$

Here $k = 2\pi/\lambda$ is the radiation wave-number, and

$$\Delta s = M_{51} x + M_{52} \theta_x + M_{56}(\Delta p / p) \qquad (21)$$

is the particle displacement on the way from pickup to kicker relative to the reference particle which experiences zero displacement and obtains zero kick, $M_{5n}$ are the elements of 6x6 transfer matrix from pickup to kicker, $x$, $\theta_x$ and $\Delta p/p$ are the particle coordinate, angle and relative momentum deviation in the pickup.

For small amplitude oscillations the horizontal and vertical cooling rates per turn are [66]:

$$\begin{bmatrix} \lambda_x \\ \lambda_s \end{bmatrix} = \frac{k\xi_0}{2} \begin{bmatrix} M_{56} - \tilde{M}_{56} \\ \tilde{M}_{56} \end{bmatrix}, \qquad (22)$$

where $\tilde{M}_{56} = M_{51} D_p + M_{52} D'_p + M_{56}$ is the partial slip-factor introduced so that for a particle without betatron oscillations and with momentum deviation $\Delta p/p$ the longitudinal displacement relative to the reference particle on the way from pickup to kicker is equal to $\tilde{M}_{56} \Delta p / p$, and $D$ and $D'$ are the dispersion and its derivative in the pickup. We assume that there is no $x$-$y$ coupling in the chicane. Introduction of $x$-$y$ coupling outside the cooling area allows a redistribution of the horizontal damping rate between two transverse planes. The sum of damping rates is: $\Sigma \lambda_n = k \xi_0 M_{56}/2$.

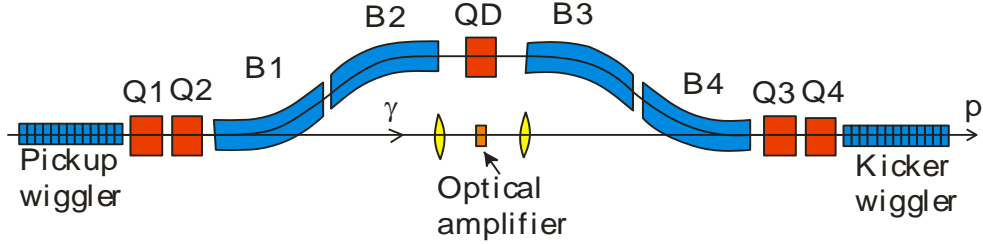

**Figure 27:** General layout of the optical stochastic cooling insert (Q- quadrupole magnet, B – dipole magnet).

Refs. [66, 67] provide detailed theoretical consideration of the OSC that takes into account major non-linearity of longitudinal motion coming from particle angles. Correction of that path length non-linearity can be achieved by two pairs of sextupoles located between the dipoles of each dipole pair of the chicane. It is also shown that the cooling dynamics is determined by a small number of parameters: the initial rms momentum spread and emittance ($\sigma_p, \varepsilon$), the wave number of optical amplifier ($k$), the dispersion invariant ($A^*$, the value of $A = (1 + \alpha^2) D^2 / \beta + 2\alpha D D' + \beta D'^2$ in the center of the chicane) and the path length delay ($\Delta s$). The value of $\Delta s$ is determined by signal delay in optical amplifier and normally should be in the range of few mm.

The IOTA OSC experiment will take one of four straight sections with length of ~4 m and will employ 2.2 μm OA and a chicane to delay the beam by $\Delta s$=2 mm – see the list of parameters in Table 8. Decreasing the ring energy from 150 to 100 MeV reduces the equilibrium emittance and momentum spread and allows sufficiently large cooling ranges to be obtained for both planes. Operation at the coupling resonance redistributes SR cooling rates between horizontal and vertical planes resulting approximately equal equilibrium transverse emittances and, consequently, twice smaller horizontal emittance. The OSC straight section optics is symmetric relative to the chicane center and is depicted in Fig. 28.

Table 8: Main Parameters of the OSC test experiment in IOTA

| Parameter | Value |
|---|---|
| Electron beam energy | 100 MeV |
| Tunes, $Q_x/Q_y$ | 6.36/2.36 |
| Transverse rms emittance, $\varepsilon_x$ | 2.6 nm |
| Rms momentum spread, $\sigma_p$ | $1.06 \cdot 10^{-4}$ |
| SR damping times (ampl.), $\tau_s/\tau_x/\tau_y$ | 1.7 / 2 / 1.1 s |
| *OSC Experiment* | |
| Radiation wavelength at zero angle | 2.2 μm |
| OSC damping times, $\tau_s/\tau_x=\tau_y$ | 0.1 /0.17 s |
| Dipole magnetic field | 2.5 kG |
| Dipole length | 8 cm |
| Horizontal beam offset | 35.1 mm |
| Undulator parameter, $K$ and period | 0.6 / 11.06 cm |
| Total undulator length, $L_w$ | 77 cm |
| Length from OA to undulator center | 1.65 m |
| Amplifier gain (power) | 5 |
| Lens focal length, $F$ | 80 mm |

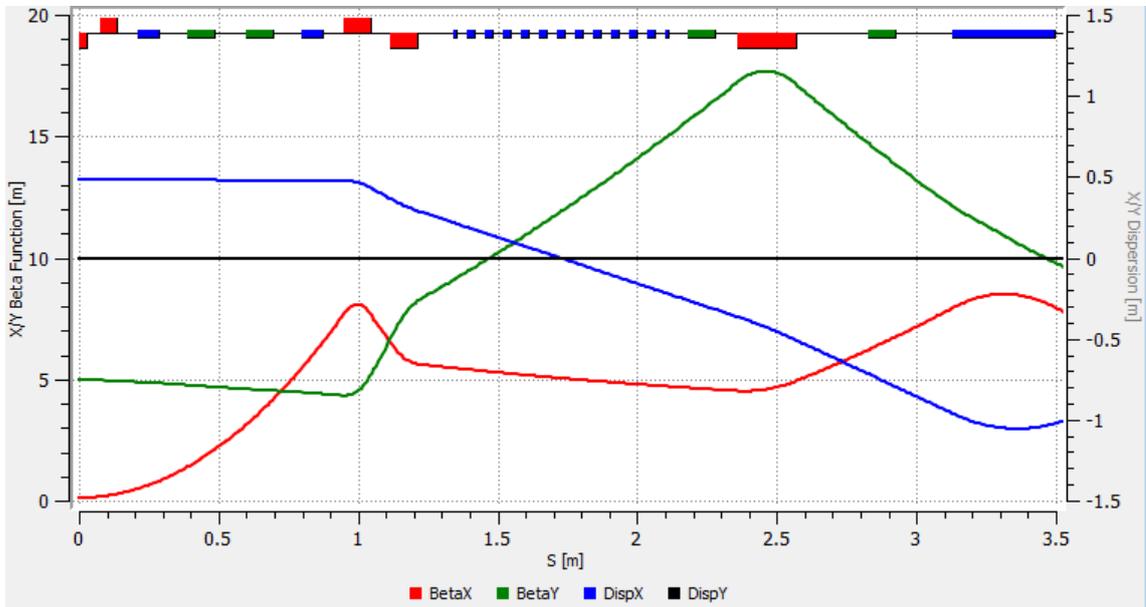

**Figure 28:** Beta-functions ($\beta_x$ – red, $\beta_y$ – green) and dispersion (dark blue) for half of the IOTA OSC straight section. The chicane center is located at $s = 0$; red rectangles at the top mark positions of quadrupoles, blue ones – dipoles and undulator.

The optical system has two lenses as shown in Fig. 27. The first lens focuses the beam radiation on the OPA crystal, and the second one focuses the amplified radiation into the kicker wiggler. The lenses have the same focal length and radius. The focal length was chosen to minimize the size of EM radiation on the OPA crystal. A longer focal length decreases the spot size at the crystal entrance but increases the waist and effects of the field of depth. An optimal value of the focal length is 8 cm with lens radius of 3.5 mm. The lens radius was determined by a compromise between the damping rates growth with the radius and an increase of signal delay leading to a drastic deterioration of beam optics. Windows in vacuum chamber located along the axis allow the pumping light to get in and out together with the amplified radiation. The radiation of pumping laser comes through the same lenses. Additional matching lenses are installed outside the vacuum chamber. The power gain of the OA is about 5. Cooling the OA to liquid nitrogen temperature is required. It increases the crystal thermal conductivity resulting in an acceptable temperature difference across the crystal (~8K) and thermal stresses. Cooling also reduces the dependence of refraction index on temperature and, thus, optics distortions related to the high pumping power.

Parameters of undulators and the cooling rates are also shown in Table 8. The undulator period was chosen so that the wavelength of the radiation at zero angle would be at the band boundary (2.2 μm). The cooling rates were computed using formulas developed in [67]. It is interesting to emphasize that even in the absence of amplification (passive system, gain equal 1) the OSC damping in IOTA will exceed the SR damping.

**3.6 Electron cooling**

Electron cooling in IOTA would extend the range of available brightness for space-charge experiments with protons. It would also provide a flow of neutral hydrogen atoms through spontaneous recombination for beam diagnostics downstream of the electron lens. Of great scientific interest is also the question of whether nonlinear integrable optics allows cooled beams to exceed the limitations of space-charge tune spreads and instabilities.

*3.6.1 Electron Cooling of Protons*

Typical IOTA proton parameters for cooling are shown in Table 9. The parameters are chosen to balance the dominant heating and cooling mechanisms, while achieving significant space-charge tune shifts. To match the proton velocity, the accelerating voltage in the electron lens has to be $V = 1.36$ kV. At these energies, proton lifetime is dominated by residual-gas scattering and by intrabeam scattering, due to emittance growth in the absence of cooling. The process of charge neutralization can also be relevant, and it is discussed below. At the residual gas pressure of $10^{-10}$ mbar, the lifetime contributions of emittance growth due to multiple Coulomb scattering and of losses from single Coulomb scattering are 40 s and 40 min, respectively. Intrabeam scattering has a stronger effect. Whereas the transverse emittance growth time is long (120 s), the longitudinal growth time can be as small as 2.5 s, indicating a possible heat transfer from the longitudinal to the transverse degrees of freedom, which must be mitigated by keeping the effective longitudinal temperature of the electrons (which is dominated by the space-charge depression and therefore by their density) low enough. At the same time, one needs to ensure that the heating term of the magnetized cooling force is negligible. One can achieve cooling rates of about 20 ms and reduce the transverse emittance by about a factor 10, with a corresponding increase in brightness.

**Table 9:** Proton beam parameters for electron cooling in IOTA.

| Parameter | Value |
|---|---|
| Kinetic energy | 2.5 MeV |
| Normalized velocity (relativistic $\beta$) | 0.073 |
| Number of particles | $5 \times 10^9$ |
| Beam current | 0.44 mA |
| Normalized rms emittance | 0.3 → 0.03 μm |
| Rms beam size at cooler | 4 → 1.3 mm |
| Relative momentum spread | $5 \times 10^{-4}$ |
| Space-charge tune shift | -0.028 → -0.28 |
| Transverse temperature (avg.) | 5 → 0.5 eV |
| Longitudinal temperature | 0.6 eV |

*3.6.2 Diagnostics through Recombination*

IOTA is a research machine and diagnostics is critical to study beam evolution over the time scales of instability growth. The baseline solution for profile measurement consists of ionization monitors, with or without gas injection. In IOTA, with $N_p = 5 \times 10^9$ circulating protons, for a residual gas pressure of $10^{-10}$ mbar, one can expect 9 ionizations per turn, or an ionization rate of 4.9 MHz.

Spontaneous recombination, $p + e^- \rightarrow H^0 + h\nu$, has proven to be a useful diagnostic for optimizing electron cooler settings and to determine the profile of the circulating beam [68]. Neutral hydrogen is formed in a distribution of excited Rydberg states, which, in order to be detected, have to survive Lorentz stripping through the electron lens toroid and through the next ring dipole. For IOTA parameters and magnetic fields, atomic states up to $n = 12$ can survive. The corresponding recombination coefficient is $\alpha_r = 9.6 \times 10^{-19}$ m³/s for $kT_e = 0.1$ eV. As a function of electron temperature $T_e$, this coefficient varies as $1/\sqrt{kT_e}$. The total recombination rate $R$ is also proportional to the fraction of the ring occupied by the cooler, $L/C = (0.7 \text{ m}) / (40 \text{ m})$ and to the electron density $n_e$:

$$R = N_p \alpha_r n_e (L/C)(1/\gamma^2). \quad (23)$$

For $N_p = 5 \times 10^9$ and $n_e = 5.8 \times 10^{14}$ m⁻³, one obtains a rate $R = 48$ kHz, which is small enough to not significantly affect beam lifetime, but large enough for relatively fast diagnostics, complementary to the ionization profile monitors.

*3.6.3 Electron Cooling and Nonlinear Integrable Optics*

A new research direction is suggested by the experimental conditions in IOTA: in the cases where electron cooling is limited by instabilities or by space-charge tune spread, does nonlinear integrable optics combined with cooling enable higher brightnesses?

It seems feasible to investigate this question experimentally in IOTA. The more straightforward scenario includes electron cooling parameters such as the ones described above. Integrability and tune spreads are provided separately by the nonlinear magnets. Space-charge tune spreads of 0.25 or more, and comparable nonlinear tune spreads, are attainable. An appealing but more challenging solution would be to combine in the same device, the electron lens, both cooling and nonlinearity (a lens of the McMillan type, for instance). If successful, such a solution would have a direct impact on existing electron coolers in machines that are flexible enough to incorporate the required linear part and phase advance of the nonlinear integrable optics scheme. Preliminary studies indicate that it is challenging to incorporate both the constraints of cooling and the high currents needed to achieve sizable tune spreads, unless one can mitigate the space-charge depression of the electron beam. This option is still under study.

As a general comment, we add that instabilities are often driven by impedance. In a research machine dedicated to high-brightness beams, it is useful to be able to vary the electromagnetic response of the beam environment. For this reason, positive feedback with a transverse damper system is being proposed to explore the stability of cooled and uncooled beams with self-fields in linear and nonlinear lattices.

**3.7 Other experimental beam studies**

Besides the experimental beam studies of relevance for the future the high energy physics intensity frontier accelerators, a few collateral experiments utilizing unique beam capabilities of the facility have been proposed or initiated by various institutional members of the IOTA/FAST collaboration. In addition to their important scientific results, these experiments are to greatly enhance training and education in beam physics and accelerator technology at the participating universities and institutions.

*3.7.1 Generation of X-rays, Gamma rays and THz radiation*

*Channeling Radiation (CR):* Electron beams, while channeling through a single crystal such as diamond, can create hard X-rays with high spectral brilliance. The intensities are large enough to be useful for medical, industrial and national security purposes. The 50 MeV electron beams at FAST can create X-rays with energies from 40 to 140 keV. The X-ray energies are tunable by changing the beam energy and are quasi-monochromatic with typical energy spreads of about 10%. Low emittance electron beams, created either by using field emitter nanotips or photocathodes with tailored laser spot sizes, can increase the spectral brilliance by orders of magnitude compared to previous CR experiments [69]. Compared to a synchrotron light sources and X-FELs, channeling in a photoinjector requires MeV scale rather than GeV scale electron beams, modest laser power for the photocathode and has a compact footprint.

When propagating parallel to a crystal plane, electrons with small enough transverse energies are trapped by the atomic potentials of the nuclei and undergo oscillatory motion and radiate. Beam channeling requires that the rms divergence be smaller than the critical angle for channeling: about 1 mrad at 50 MeV. For beam energies below 100 MeV, the CR spectrum is discrete and best understood as a quantum mechanical process. At the beam parameters of FAST, we expect a photon yield of about $10^{-4}$ photons per electron. A detailed theoretical model of the expected CR spectrum at FAST, discussion of dechanneling and rechanneling and a comparison with previous CR results from the ELBE linac can be found in [70]. The main source of crystal induced background to the CR is incoherent bremsstrahlung. Like CR, this co-propagates with the electron beam but has a larger angular spread than the typical *1/γ* cone of CR. Past experiments with electron energies below 30 MeV have reported ratios of the CR peak to the BS background in the range from 6 to 10. Fig. 29 shows the expected CR spectrum with 43 MeV electrons under three different assumptions for the channeling yield, compared to the expected BS background [71]. There are CR lines at (51, 68, 110) keV and the ratio of the CR peak at 110 keV to BS ranges from 4 to 15 in this calculation.

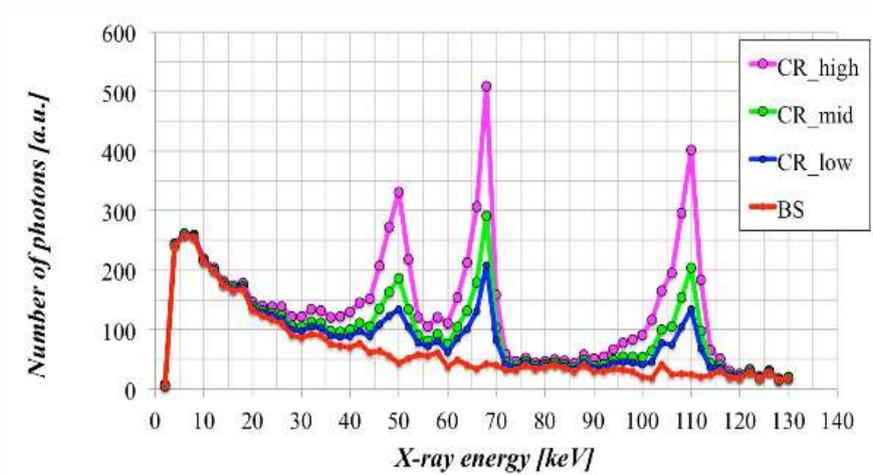

**Figure 29**: Expected channeling radiation spectra with a 43 MeV electron beam. Also shown is the background from bremsstrahlung (BS).

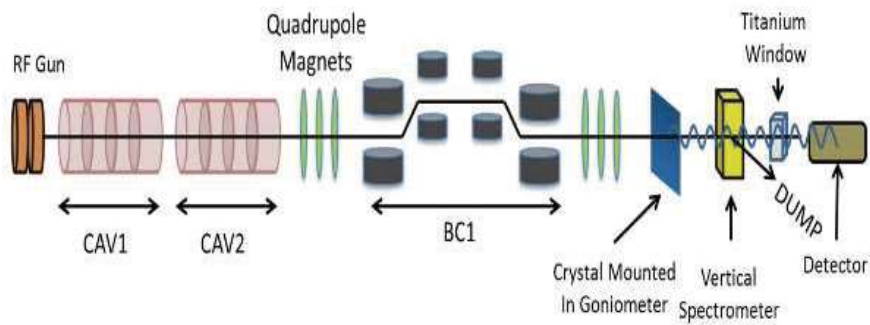

**Figure 30**: Schematic layout of the channeling radiation experiment in the IOTA/FAST electron injector beamline. CAV1,2 – SRF capture cavities, BC1- magnetic chicane bunch compressor. Not shown here are a polycarbonate plate inserted in front of the forward detector for Compton scattering the X-rays and another X-ray detector at 90º to the beamline to detect the Compton scattered X-rays.

Beam commissioning and the first results of the CR experiment at FAST (see layout in Fig. 30) are reported in [72]. Preventing pile up and saturation in the single photon counting detectors requires operating with very low charge, less than 100fC for the forward detector and less than 200pC for the orthogonal detector. The primary challenge during this study was to mitigate the dark current (from the RF gun) induced X-rays. The measures to reduce the dark current included lowering the voltage on the RF gun, optimizing the solenoid currents, inserting a dark current collimator and finally scraping the off-energy dark current in the chicane. During the next run, the dark current induced background will be substantially further reduced by relocating the detectors and more shielding will be added.

*Parametric X-rays (PXR):* Another mechanism to produce X-rays is via the Bragg reflection of virtual photons (accompanying the electrons) off specific crystal planes. Compared to channeling, the X-ray energy is independent of the beam energy, and the photon energies can be tuned by rotating the crystal. In addition, the X-rays are emitted at large angles from the beam direction, hence are freer from the bremsstrahlung background and have a narrower line width of about 1%. The PXR photons can also be emitted simultaneously with channeling radiation. At FAST, X-rays with energies in the range from 3-18 keV can be emitted depending on the plane of reflection. A detailed study characterizing PXR production in FAST can be found in [73].

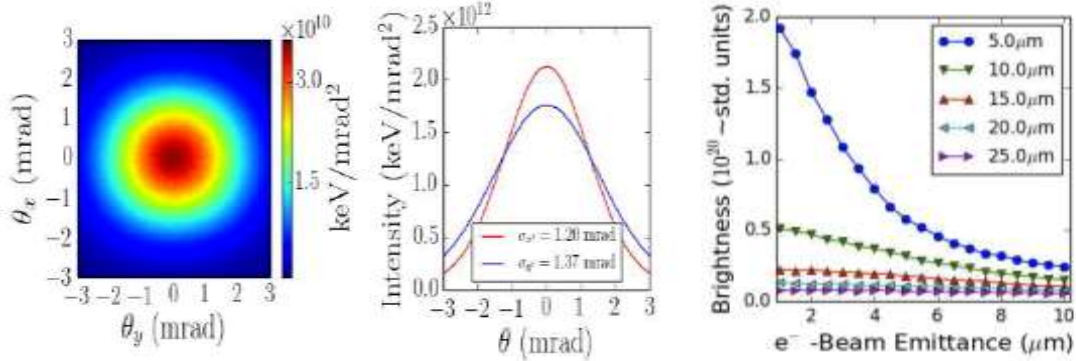

**Figure 31**: Example of simulated gamma-ray angular distribution (left) with associated projection (middle) and gamma-ray brightness as function of electron beam emittance for different waist size at the electron-laser interaction point (right).

*Gamma rays from Inverse Compton Scattering:* The electron beam energy in the IOTA/FAST beamline will be increased to 300 MeV when the accelerating cryomodule (CM) becomes operational. This will allow the generation of hard X-rays or gamma rays by scattering the electron beam off laser photons via the process generally called Inverse Compton Scattering (ICS). The electrons oscillate in the transverse electric fields of the laser pulse and emit radiation in the direction of their motion. For head-on collisions between the electron and laser beams, the energy of the photons emitted along the electron beam axis is related to the laser photon energy $E_L$ as $E_\gamma = 4\gamma^2 E_L$. Thus, a 1eV laser photon and a 300 MeV electron beam, can create ~1.5 MeV photons via the ICS process. High photon rates and spectral brilliance require the use of a high power laser system. In the IOTA/FAST injector, the same IR laser system used to generate the electrons from the photocathode will be used to collide with the electron beam. The laser beam energy will in the first stage be increased from 50 μJ to about 50 mJ by two single pass amplifiers. In the second stage, a high finesse optical cavity will be used for amplification to raise the laser beam energy to the Joule level. The scheme is to trap each laser pulse in the cavity with low loss mirrors at each end and coherently combine several laser pulses before colliding them with an electron bunch in the cavity [74]. Both laser and electron beams will have to be focused to small spot sizes at the collision point to increase the number of scattered photons produced while high brightness also requires a small electron bunch length. With typical FAST parameters, we expect to produce about $10^6$ photons per electron bunch and a spectral brilliance of about $10^{20}$ photons/s-(mm-mrad)$^2$ – 0.1% BW and a photon energy spread of about 1% (Fig. 31). Design of the optical cavity with high amplification, while controlling position and angular misalignments within allowed limits, and synchronization of the colliding pulses will be the major issues. Progress in the construction and operation of a four mirror optical cavity was recently reported [75]. A higher laser sampling rate above the present value of 3MHz will also help in reducing the cavity length and will be explored as a possible option.

*Tunable THz generation:* A scheme has been proposed [76] to generate THz radiation using MeV scale beams moving through a slit mask in the BC1 chicane. If the beam enters the chicane with an energy chirp, the transversely separated beam fragments after the mask are longitudinally separated at the ends of the chicane into a train of sub-picosecond-duration bunches. When these bunches impinge on an *Al* foil, they generate coherent THz radiation via coherent transition radiation [77]. If the emittance is lowered in the horizontal plane prior to entering the chicane, e.g. by a flat beam transformation [78], the THz radiation spectrum can

– 34 –

extend from 1-4 THz, the tuning can be done by proper choice of the slit mask width and the RF chirp.

*3.7.2 Opportunities for advanced beam dynamics studies*

Over the last decade Fermilab has pioneered the experimental development of phase space manipulations between two degrees of freedom such as the round-to-flat beam transformation and the transverse-to-longitudinal phase-space exchange [78, 79, 80]. Some of these transformations will be further investigated at FAST and possibly combined with applications such as the one discussed in the previous sub-section.

The FAST photoinjector was designed to enable the generation of magnetized beams formed by applying an axial magnetic field on the photocathode. Such a magnetized beam when decoupled using a skew-quadrupole-magnet channel can be transformed into a beam with a large transverse emittance ratio. The required skew quadrupole magnet channel was integrated into the 50-MeV beamline. Numerical simulations have shown that such a setup will enable the generation of flat beams with transverse emittance ratio close to 500 [81]. In addition, the flat beam transformation is located upstream from the magnetic bunch compressor (BC1 in Fig. 30) thereby enabling the production of high-peak-current flat beams when compressed in BC1 [82]. Such beams could have applications in the development of Smith-Purcell free-electron lasers [83], beam-driven advanced acceleration techniques [84] or support tests of short-period undulator magnets [85]. Likewise, the production of magnetized beams has relevance to electron cooling of hadron beams [86]. Exploring how magnetized beams transform (e.g. via decompression in BC1) could help validate some of the concepts for electron cooling in the Electron-Ion Collider (EIC). Additionally, the injection of magnetized beams in IOTA could support the investigation of collective effects (especially micro-bunching instabilities induced via longitudinal space charge and coherent synchrotron radiation) that could hamper the performances of multi-pass electron cooling.

As detailed in Section 2.2 above, the IOTA/FAST photoinjector incorporates a versatile photocathode laser system. The UV laser pulse can especially be manipulated using alpha-BBO crystal to produce shaped temporal profiles [87]. Preliminary use of this capability has included the formation and acceleration of double-bunch MeV beams [88] which could be used to probe the longitudinal beam dynamics and possibly investigate collective effects by extending a technique developed in [89]. Likewise, expanding the technique to produce a train of bunches with picosecond separation could drive longitudinal-space charge-driven micro-bunching instability [90] which, in turn, could support the production of coherently-enhanced THz radiation downstream of BC1 as a complementary way to the method described above.

Another important topic of research to be enabled at FAST is the development of laser- or radiation-based electron beam control methods. The planned optical-stochastic cooling at IOTA is an example of such research. In the FAST injector system, relevant to the OSC (see Section 3.5), could be tested, for instance several undulator magnets currently under construction [91] to produce the necessary wavelength (~2 µm) to test a single-pass Cr:ZnSe amplifier currently under development for the OSC experiment [92]. Additionally, two undulator magnets separated with the proper magnetic lattice could enable the exploration of information preservation, laser-assisted beam conditioning methods [93], or the generation of gamma rays ICS backscattered as the undulator radiation produced by a bunch interacts with a subsequent electron bunch.

Finally, the coupling of a photoinjector with an SRF linear accelerator is foreseen to support research related to beam-quality preservation during acceleration in SRF accelerating cavities. An ongoing work relates to the measurement of the transverse transfer matrix of the TESLA cavity composing the cryomodule [94]. This study could be further expanded by using a flat beam as a probe to investigate emittance dilution of the smaller emittance. Additionally, the photoinjector electron beam could probe long-range wakefield effects which could affect the multi-bunch beam parameters. Although similar investigations have been performed for the case of regularly spaced bunches within a macropulse, the versatility of the FAST laser system will enable these investigations to be expanded to the case of irregularly spaced bunch patterns. Flexible bunch patterns within a macropulse are foreseen to be employed in the MARIE project currently under consideration at Los Alamos National Laboratory [95].

**4. Status and Plans**

A staged approach is being used for installation and commissioning of the IOTA/FAST facility, with each of three major sections being broken into parts to allow for testing. These parts consist of the electron injector, the proton injector, and the IOTA ring.

The electron injector beamline was originally intended as a 1% test of SRF technology for the International Linear Collider (ILC) [96]. The construction of this facility is a large undertaking, which is being done in multiple phases. The first phase included the installation of the infrastructure necessary to operate the facility: cryogenics, water, power, RF, and controls, as well as building the test cave. Subsequent phases included the installation and operation (at 2 K) of the SRF cryomodules, CC1, CC2 (each containing a single 9-cell cavity) and CM2 (containing eight 9-cell cavities). In 2011 a large construction project was completed which added a 70 m-long tunnel to the existing FAST facility. This expansion essentially doubled the length of the test accelerator from 75 m (the length of the existing NML building) to 140 m. In addition to providing additional length to the overall accelerator, the expansion also includes a large 15 m-wide area for the high-energy test beam lines at the downstream end of the accelerator; as well as space to install the 40 m-circumference IOTA storage ring to conduct future accelerator R&D programs and experiments. An enclosure to house the high-energy beam absorbers and electrical service building were also constructed at the end of the test beam lines.

First parts of the SRF electron accelerator were delivered and tested in 2008, but the accelerator only took shape with the purpose of delivering electrons to the IOTA source in 2011-14. In 2015, a 20 MeV low-energy electron beam run was performed through the low-energy beamline using only the 1.3 GHz, 1.5-cell, normally conducting gun at 5 MeV and a single 1.3 GHz, 9-cell, TESLA-style, SRF cavity. Once complete, construction of building infrastructure continued while a second SRF cavity was completed and installed, allowing for a full 50-MeV electron beam run in 2016. Between the two low-energy commissioning periods, a number of beamline parameters were found as shown in the Table 9.

**Table 9:** Summary of primary IOTA/FAST low-energy electron injector beam parameters. Parameters verified previously, but not as part of the 50-MeV run are denoted with an asterisk.

| *Parameter* | *50 MeV IOTA electron injector value* |
|---|---|
| Beam energy | 20 MeV – 50 MeV |
| Bunch charge | < 10 fC – 3.2 nC per pulse |
| Bunch train (Macropulse) | 0.5 – 9* MHz for up to 1 ms (3 MHz nominal) |
| Train frequency | 1 – 5* Hz |
| Bunch length | Range: 0.9 – 70* ps (5 ps nominal) |
| Bunch emittance (50 pC/pulse) | Horz: $1.6 \pm 0.2$ μm    Vert: $3.4 \pm 0.1$ μm |

In the low-energy beam commissioning runs, no beam was delivered to the 12-m long 1.3 GHz, TESLA-style, SRF cryomodule, which was conditioned with RF to a level of 31.5 MeV/m average gradient for all eight cavities within the cryomodule [97]. The high energy section of the electron accelerator is being built in the newly completed accelerator enclosure ("cave") extension. It will initially transport beam from the cryomodule to the high-energy beam absorber in early 2017. The next phase of the project includes the installation and commissioning of the high-energy portion of FAST, and the acceleration and transport of 300 MeV electron beam to the high-energy beam absorber. This phase involves extending the existing test cave and all associated utilities to tie-in with the newly constructed tunnel expansion, as well as the installation of the accelerator components (magnets, supports, vacuum systems, instrumentation, etc.) that comprise the 300 MeV beamline of FAST. Construction of the 300 MeV beamline began in the summer of 2016 and is expected to be complete and ready for beam commissioning in mid-2017.

In parallel, a proton source for the IOTA ring is being refurbished from the former High Intensity Neutrino Source (HINS). The proton source itself is a duo-plasmatron that feeds an RFQ before the 2.5 MeV proton ($p^+$) beam passes through a debuncher cavity. The original HINS source included a number of other RF structures which will not be used, and the cooling problems that prevented the RFQ from reaching its originally specified duty factor will not be an issue for this application. Recommissioning of the 2.5 MeV proton ($p^+$) beam is expected from the RFQ in 2017. The beam will be characterized using an Allison Scanner, Faraday cup, and other instrumentation in situ at its current location, the MDB building. This will to allow the RFQ commissioning to progress independently from any construction or commissioning constraints with respect to the electron injector or the IOTA ring. Once the RFQ proton source commissioning is complete at MDB, components will be moved from MDB to the FAST facility at NML to be reassembled alongside the high energy beamline in the NML cave extension, with commissioning of the 2.5 MeV proton ($p^+$) beam into the IOTA ring in 2019.

Finally, the construction of the IOTA ring is also underway. Installation of the ring components began in 2016 and is expected to be complete and ready for initial commissioning with 150 MeV electrons from the FAST accelerator in late 2017 or early 2018. Beyond this, all configuration changes are expected to be related to installation of apparatus or components of various beam physics experiments, including implementation of an inverse Compton scattering (ICS) experiment in the high energy electron beamline, nonlinear magnets for integrable optics, the electron lens system and elements of the optical stochastic cooling system in the IOTA ring.

## 5. Summary

We have presented the design and main parameters of the Integrable Optics Test Accelerator (IOTA) facility at Fermilab, which is set to become be the only accelerator R&D facility in the US (and in the world, for that matter) that supports development of new ideas towards the next generation high-intensity proton facilities and allows a broad range of intensity-frontier-motivated experiments, such as *integrable optics* with non-linear magnets and with electron lenses, and *space-charge compensation* with electron lenses and electron columns, as well as on many other interesting research topics. We have described in detail the main components of the facility, including the storage ring for advanced beam physics research and its 70 – 150 MeV/c proton and electron injectors, and outlined the facility construction status and plans, as well as physical principles, design, and hardware implementation plans for the experiments planned.